\documentclass{aa}
\usepackage{graphicx}
\usepackage{natbib}
\bibpunct{(}{)}{;}{a}{}{,} 
\usepackage{txfonts}
\begin{document}
   \title{Dynamics of the solar atmosphere above a pore with a light bridge}
   
   \author{M. Sobotka
          \inst{1}
          \and
          M. \v{S}vanda
          \inst{1,2}
          \and
          J. Jur\v{c}\'{a}k
          \inst{1}
          \and
          P. Heinzel
          \inst{1}
           \and
          D. Del Moro
          \inst{3}
          \and
          F. Berrilli
          \inst{3}
         }

   \offprints{M. Sobotka}

   \institute{Astronomical Institute, Academy of Sciences of the Czech Republic
              (v.v.i.), Fri\v cova 298, CZ-25165 Ond\v rejov, Czech Republic\\
              \email{msobotka@asu.cas.cz}
         \and
             Charles University in Prague, Faculty of Mathematics and Physics,
             Astronomical Institute, V Hole\v{s}ovi\v{c}k\'{a}ch 2,
             CZ-18000 Praha 8, Czech Republic
         \and
             Department of Physics, University of Roma Tor Vergata, Via della
             Ricerca Scientifica 1, I-00133 Roma, Italy\\
             }

   \date{Received June 26, 2013; accepted September 30, 2013.}
   
   \abstract
   {Solar pores are small sunspots lacking a penumbra that have a prevailing
vertical magnetic field component. They can include light bridges
at places with locally reduced magnetic field.
Like sunspots, they exhibit a wide range of oscillatory phenomena.}
   {A large isolated pore with a light bridge (NOAA 11005) is
studied to obtain characteristics of a chromospheric filamentary
structure around the pore, to analyse oscillations and waves
in and around the pore, and to understand the structure and brightness
of the light bridge.}
   {Spectral imaging observations in the line Ca~II 854.2~nm and
complementary spectropolarimetry in Fe~I lines, obtained with the
DST/IBIS spectrometer and HINODE/SOT spectropolarimeter,
were used to measure photospheric and chromospheric velocity fields,
oscillations, waves, the magnetic field in the photosphere, and acoustic
energy flux and radiative losses in the chromosphere.}
   {The chromospheric filamentary structure around the pore has
all important characteristics of a superpenumbra: it shows an
inverse Evershed effect and running waves, and has a similar morphology
and oscillation character. The granular structure of the light
bridge in the upper photosphere can be explained by radiative heating.
Acoustic waves leaking up from the photosphere along the inclined
magnetic field in the light bridge transfer enough energy flux to balance
the total radiative losses of the light-bridge chromosphere.}
   {The presence of a penumbra is not a necessary condition for
the formation of a superpenumbra. The light bridge is heated by radiation
in the photosphere and by acoustic waves in the chromosphere.}

   \titlerunning{Chromosphere and photosphere of a solar pore}

   \maketitle
%

\section{Introduction}\label{sec:intro}

Pores are small sunspots without a penumbra. The size of the smallest
pores is similar to the size of a granule, while the largest pores
exceed the size of small sunspots with penumbrae. Pores are generally
darker than the intergranular lanes, but are somewhat brighter than the
umbrae of large sunspots \citep{TW:2008}. The absence of a filamentary
penumbra in the photosphere of pores has been interpreted as an
indication of a simple magnetic structure with a mostly vertical field
\citep[e.g.][]{SimonWeiss:1970,Rucklidge:1995}. Magnetic-field lines
are observed to be nearly vertical in the centres of pores and are
inclined to the local normal by about 40\degr ~to 60\degr ~at their edges
\citep{kepmar:96,Suetter:98}.
Pores contain a wide variety of fine bright features, such as umbral dots
and light bridges, which may be signs of a convective energy transportation
mechanism \citep[e.g.][]{sobetal:99,keiletal:1999,hirzberg:03,giordano:08,
sobjur:09, Ortizetal:2010}.
The simple magnetic structure of pores provides an opportunity to study
the interaction of plasma with the magnetic field. Relations between
the magnetic field and the velocity fields in the photosphere of a large
pore and its surroundings were described in our previous paper
\citep{SobMorEtal:2012}. In this work we aim to obtain characteristics
of a chromospheric filamentary structure observed around the pore, to
analyse oscillations and waves, and to understand the structure and
brightness of a light bridge inside the pore.

Light bridges (hereafter LBs) are bright structures in sunspots and pores
that separate umbral cores or are embedded in the umbra. Their
structure depends on the inclination of the local magnetic field and can be
granular, filamentary, or a combination of both \citep{sobotka:1999}.
Widths of LBs vary from smaller than 1\arcsec ~to several seconds of arc
and the brightnesses range from the intensity of umbral dots to photospheric
brightness. Some LBs show a long narrow central dark lane running
along the length of the bridge \citep{BeBe:2003}. Many observations
confirm that the magnetic field in LBs is generally weaker and more inclined
with respect to the local normal \citep{BeckSchr:1969,Lites:1991,
Ruedi:1995,Leka:1997}. \citet{JurLB:2006} showed that in LBs the
field strength increases and the inclination decreases with increasing
height. This indicates the presence of a magnetic canopy above a field-free
region located in deep layers that intrudes into the umbra and
forms the LB. Convective elements similar to granulation with upflows
in bright granules and downflows in dark lanes are observed in granular
LBs \citep{Rimmele:1997}.
Above LB, \citet{BeBe:2003} found a persistent brightening in the
TRACE 160 nm bandpass formed in the chromosphere. The brightness
reaches the level of the magnetic plage surrounding the sunspot.
These authors interpreted this as a steady-state heating possibly due to
constant small-scale reconnections in the inclined magnetic field in the
LB chromosphere. This opinion was also supported by \citet{Shimi:2012}.
An alternative source of heating of the LB chromosphere may be the
energy transferred by acoustic waves, as will be shown in this paper.

In the chromosphere, large isolated sunspots are often surrounded
by a pattern of dark, nearly radial fibrils. This pattern, called
superpenumbra, is similar to the white-light penumbra, but extends
to a much larger distance from the sunspot. Dark superpenumbral
fibrils typically begin near the outer edge of the penumbra, but
about a third of them begins well within the penumbra, sometimes
near the umbral border \citep{TW:2008}. The dark fibrils are often
slightly curved and the spacing between them increases from
1\arcsec ~near the umbra to 2\arcsec--3\arcsec ~at the outer edge of
the superpenumbra. The superpenumbra hosts the inverse
Evershed flow -- an inflow towards the sunspot in the chromosphere,
combined with a downflow near the umbral border.
Time-averaged Doppler measurements indicate that the highest speed of
this flow is equal to 2--3~km~s$^{-1}$ near the outer penumbral border
\citep[e.g.][]{Aliss:1988}. Recently, \citet{Vissers:2012} reported
high-speed intermittent (flocculent) flows, observed in the form of
features hosting Doppler velocities of $\pm 20$ km~s$^{-1}$ that
moved along superpenumbral and plage-related fibrils seen
in the H$\alpha$ line centre.
An open question is still whether the superpenumbra can be formed
also around pores, without a need for a proper penumbra.

Sunspots and pores exhibit a wide range of oscillatory phenomena.
They were studied since the discovery of umbral flashes by
\citet{BeckTal:1969}. Umbral flashes are reoccurring brightenings
in strong Ca~II lines with periodicity of roughly three minutes.
Later, umbral oscillations with dominant periods in the five-minute
band (3--4 mHz) in the photosphere \citep{Bhatnagar:1972,Lites:1992}
and in the three-minute band (5--6 mHz) in the chromosphere
\citep{BeckSchultz:1972,Brynild:1999} were found and have been
extensively studied since
\citep[see][for a review and more references]{Bogdan:2006}.
Running penumbral waves \citep{Giovanelli:1972,Zirin:1972} are
disturbances propagating radially in the form of dark and bright arcs
from the inner to the outer penumbral boundaries. Their phase speed
is around 10--15 km~s$^{-1}$ and they have a frequency in the five-minute
band \citep{Christop:2000,Tzio:2006}. In the H$\alpha$ line centre the
waves appear to propagate beyond the outer edge of the penumbra
to the superpenumbra.
Distinct morphological parts of the evolved sunspot (umbra,
penumbra, superpenubra) seem to have different seismic properties
\citep{tziotzio2007}. The oscillation power spectrum and the
running penumbral waves might possibly be used to indicate the
presence of a superpenumbra also in the case of pores.

Acoustic flux carried by oscillations and waves to the higher
atmosphere is one possible phenomenon responsible for the heating
of the solar atmosphere \citep{Klimchuk:2006}. The role of oscillations
in the heating of the quiet-Sun chromosphere is most likely negligible,
but in  magnetic regions this may be completely different because of
the conversions  of pressure waves from below the photosphere
to magneto-acoustic modes and Alfv\'en waves \citep[e.g.][]{KhoCall:2012},
which may propagate upwards and carry a significant energy flux.

In this paper we focus on observations of a large solar pore with
LB in the infrared line Ca~II 854.2 nm. The wings of
this line are formed in the photosphere, while its core comes from
the middle chromosphere. The formation height quickly shifts from
the photosphere to the chromosphere at the transition between the
wings and core thanks to a gap in the line opacity at the temperature
minimum, as described by \citet{Cauzzi:2008}. According to Fig.~5 in
their work, the wings that are 50--70 pm away from the line centre map
the middle photosphere around $h = 200$--300 km above the unity optical
depth in the 500 nm continuum ($\tau_{500} = 1$) and the core is formed
in the height range $h = 900$--1400 km. This provides a good tool for
studying the pore and its surroundings at different heights in the
atmosphere.

The paper is organised as follows:
In Section~\ref{sec:obs}, we describe the observations and general
characteristics of the pore.
The data processing and analysis are described in Section~\ref{sec:proc}.
In Section~\ref{sec:filam}, we focus on a chromospheric filamentary
structure around the pore.
Oscillations and waves are analysed in Section~\ref{sec:osc}.
In Section~\ref{sec:LB} we study the structure of the LB and discuss
the possibility that its chromospheric layers are heated by acoustic waves.
Discussion and concluding remarks are presented in Section~\ref{sec:disc}.

\section{Observations}\label{sec:obs}

A large isolated solar pore (NOAA 11005) was observed on 15 October
2008 from 16:34 to 17:43 UT with the Interferometric Bidimensional
Spectrometer \citep[IBIS,][]{Cavallini:2006} attached to the Dunn
Solar Telescope (DST).
During the observations, the seeing conditions varied from good to
fair, but the high contrast of the pore allowed us to use it as a
lock-point for the wavefront sensor. Therefore, the adaptive optics
(AO) system of the NSO/DST \citep{Rimmele:2004} performed well
in terms of stability and resolution in the centre of the field of
view (FOV).

The active region NOAA 11005 was present on the solar disc from
11 to 16 October 2008. The slowly decaying pore, located at 25.2 N
and 10.0 W (heliocentric angle $\vartheta = 23$\degr) during our
observation, led a bipolar active region, in which the
following magnetic polarity was too weak to produce sunspots or pores.
A strong granular LB separated the pore into two umbral cores;
the larger one was in the east. The effective diameter of the pore
calculated from its area including the LB was 8\farcs 5 (6200 km).
The highest magnetic-field strength $B$ was found in the eastern
core. Its value increased with time from 1900 G to a maximum
of 2100 G. The magnetic field of the pore was, as a whole, tilted
to the west by 10\degr~with respect to the vertical. After removing
this overall tilt, the average magnetic field inclination at the
pore's edge was 40\degr~\citep{SobMorEtal:2012}.

The IBIS dataset consists of 80 sequences, each containing
a full Stokes ($I, Q, U, V$) 21-point scan of the Fe~I 617.33 nm line
and a 21-point $I$-scan of the Ca~II 854.2 nm line. The wavelength
distance between the spectral points for the Fe I line is 2.0 pm
and 6.0 pm for the Ca~II line.
The spectropolarimetric data acquisition strategy we used
was to acquire six modulation states $I \pm [Q, U, V]$ at each
wavelength position in the Fe~I line. Each sequence therefore consists
of $21 \times  6$ (Fe I) $ +\ 21$ (Ca II) = 147 narrowband images.
The exposure time for each image was set to 80~ms and each spectral
scan took 52 s to complete, thus setting the time resolution.
From this period, the time needed to scan the $I$-profile of the
Ca~II line was only 6.4 s.
The pixel scale of these $512 \times 512$ pixel images was 0\farcs 167.
Due to the spectropolarimetric setup of IBIS, the working FOV was
$228 \times 428$ pixels, that is, 38\arcsec $\times$ 71\farcs 5, with
the pore located in the centre.

For each spectral image, we also acquired a broad-band (WL) and
a G-band counterpart, both imaging approximately the same FOV.
The G-band images were not used in this work.
The pixel scale of the $1024 \times 1024$ pixel WL images ($621.3 \pm 5$ nm)
was 0\farcs 0835 and the acquisition time was 80 ms, using a shared
shutter with IBIS to ensure strictly simultaneous exposures of the WL
and spectral images.

The spectro-polarimetric data obtained in the line Fe~I 617.33 nm
were analysed by \citet{SobMorEtal:2012}. At this wavelength, the
authors estimated the spatial resolution to be about 0\farcs4,
the spectral stray-light contamination 1--2 \%, and the spatial
stray-light contamination 15 \%.
The instrumental characteristics of IBIS for the Ca~II 854.2 nm line
were described by \citet{Cauzzi:2008}. The {\it FWHM} of the
instrumental spectral transmission at 854 nm is 4.4 pm \citep{ReCa:2008},
the spectral stray-light contamination is 1.5 \%.

Complementary observations were obtained with the HINODE/SOT
Spectropolarimeter \citep{Kosugi:2007,Tsuneta:2008}. The satellite
observed the pore on 15 October 2008 at 13:20 UT, that is, 3 hours
14 minutes before the start of our observation. From one spatial scan
in the full-Stokes profiles of the lines Fe~I 630.15 and 630.25 nm we
used the part covering the pore. The pixel scale of this observation
was 0\farcs 3.

\section{Data processing and analysis}\label{sec:proc}

For each of the 80 sequences in the dataset we obtained three multi-frame
blind deconvolution \citep[MFBD,][]{Noort:2006} restored images,
each computed from 49 original WL images.
The best restored WL image in each sequence (with the highest contrast)
was used as a reference to align and de-stretch the original WL images.
The obtained correction maps were applied to the narrow-band Fe~I and
Ca~II images via the de-stretching technique to minimise the degradation
of the spectral scans caused by image motion and distortion. Although
the wavelength difference between the WL and Ca~II spectral images was
large, the spectral scans were de-stretched satisfactorily, without any
evident artefact, and the image quality benefited greatly from the process.
Then, we applied the standard reduction pipeline \citep{viticchie2010}
on the spectral data, correcting for the instrumental blueshift
\citep{Cavallini:2006} and the instrument- and telescope-induced
polarisations.

The observations in the Ca~II 854.2 nm line are strongly influenced
by oscillations and waves. Particularly the line core shows significant
shifts in wavelength due to chromospheric oscillations. The observed
intensity fluctuations in time are caused by real changes of intensity
as well as by Doppler shifts of the line profile.
To separate the two effects, Doppler shifts of the line profile were
measured using the double-slit method \citep{Garcia:2010}, consisting
in the minimisation of difference between intensities of light passing
through two fixed slits in the opposite wings of the line. This
method, formerly used in scanning photoelectric magnetographs to measure
line-of-sight (LOS) velocities, is very reliable but it does not take into
account the asymmetry of the line profile, for instance, the changes of
the LOS velocity with height in the atmosphere.

An algorithm based on this principle was applied to the time sequence
of Ca~II profiles. The distance of the two wavelength points (\emph{slits})
was 36~pm, so that the \emph{slits} were located in the inner wings near
the line centre, where the intensity gradient of the profile is at
maximum and the effective formation height in the atmosphere is
approximately 1000 km. The wavelength sampling was increased by
the factor of 40 using linear interpolation, thus obtaining the
sensitivity of the Doppler velocity measurement 53 m~s$^{-1}$.
The reference zero of the Doppler velocity was defined as a time- and
space average of all measurements. This way we obtained a series of
80 Doppler velocity maps of the working FOV 38\arcsec $\times$ 71\farcs 5.
An alternative measurement of Doppler shifts based on a parabolic fit
of the central part of the Ca~II profile in the range $\pm$ 18~pm
gave practically identical results, with only slightly (4 \%) higher
amplitudes.

Using the information about the Doppler shifts obtained with the
double-slit method, all Ca~II
profiles were shifted to a uniform position with subpixel accuracy.
This way we obtained an intensity data cube ($x, y, \lambda$, scan),
where the line centre was set to a fixed position and the spectral
range was from $-66$ pm to $+54$ pm with respect to the line centre.
If we neglect line asymmetries and inaccuracies of the method,
the observed intensity fluctuations now correspond to the real intensity
changes. The residual crosstalk between the intensity and Doppler
signals due to errors of the line-shift compensation can be estimated
from the difference of intensities in the blue and red wings at
$\pm$ 6--30 pm from the line centre, assuming symmetry of the line.
For each wavelength in the range, the normalised differences
$2\,{\rm (red-blue)/(red+blue)}$ are smaller than $\pm$ 4 \%, so that
the crosstalk between the intensity and Doppler signals is sufficiently
small.

The oscillations and waves were separated from the slowly evolving
intensity and Doppler structures by means of a $k$--$\omega$
Fourier filter. The oscillatory phenomena disappeared in the filtered
time-series of images when the phase-velocity cutoff was set to
6 km s$^{-1}$.
The unfiltered data were used to study oscillations, waves, and umbral
flashes, while the filtered data provided the information on
morphology, dynamics, and evolution of chromospheric fibrils, LB, and
other structures.
The oscillations in various regions in the FOV and at various
heights in the atmosphere (given by the position in the spectral line)
were studied using standard Fourier analysis (see Section~\ref{sec:osc}).
Given the length of the time-series and the sampling rate, the frequency
resolution is $0.24$~mHz. This is fine enough to investigate changes in
the oscillation power-spectra within the two umbral cores,
the LB, the filamentary region around the pore, and also in the
quiet-Sun region for comparison.

We used the Stokes inversion code based on response functions \citep[SIR,][]{RuizCobo:1992} to retrieve plasma properties in
the photosphere. This code works under the assumption of local
thermodynamical equilibrium and hydrostatic equilibrium. It starts
with an initial model atmosphere that is modified in several node
points for each inverted plasma parameter to obtain the best fit
of the synthetic Stokes profiles to the observed ones.
The nodes correspond to certain heights in the atmosphere and their
number determines the number of free parameters. The details of the
Fe~I 617.3 nm inversion process and obtained results were described
by \citet{SobMorEtal:2012}. Since the spectropolarimetric information
came from only one spectral line, it was necessary to keep the number
of free parameters as low as possible, allowing for only two nodes in
the temperature and one node in all other parameters such as the
magnetic-field vector and the LOS velocity.

To retrieve a more realistic distribution of plasma parameters with
height in the photosphere of the pore and LB, we calculated new inversions
using the complementary HINODE observations (see Section~\ref{sec:obs})
in two lines Fe~I 630.15 and 630.25 nm. Four nodes were set for the
temperature and three nodes for the magnetic-field strength, inclination,
and LOS velocity. Magnetic-field azimuth and microturbulent velocity were
kept constant with height. The macroturbulent velocity was set to zero.
We assumed a one-component model of the atmosphere and did not account
for stray light.

\begin{figure*}
  \centering
  \includegraphics[width=18cm]{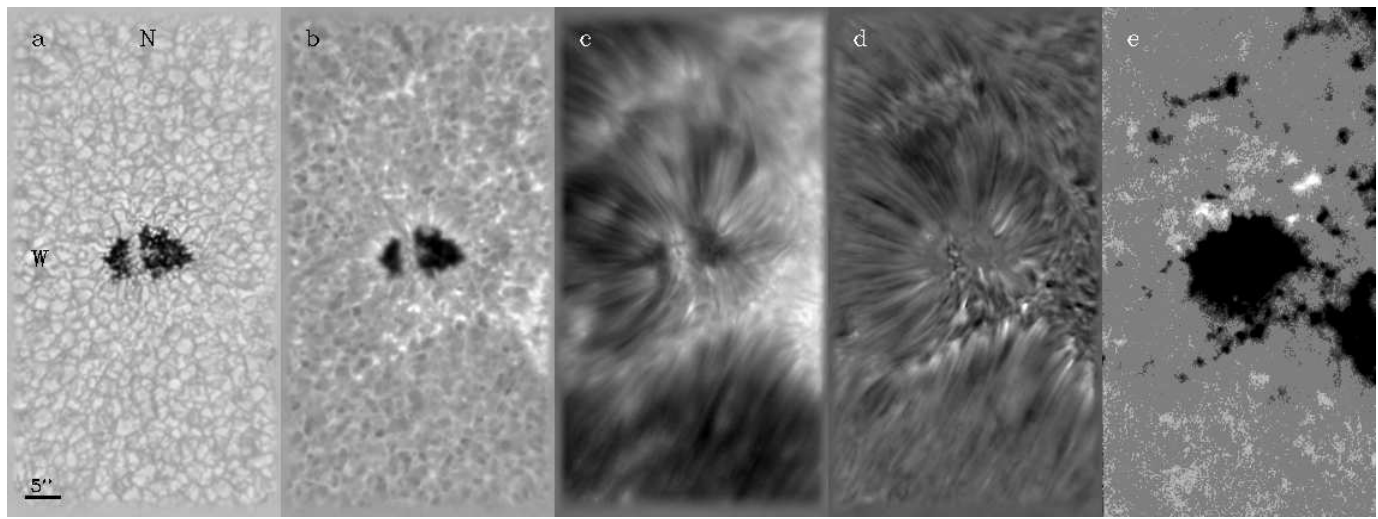}
  \caption[]{\label{fig:exa}
Pore NOAA 11005, working FOV 38\arcsec $\times$ 71\farcs 5.
North is at the top, west to the left. From left to right: (a) broad-band
image around 621~nm, (b) Ca~II 854.2 nm blue ($-60$ pm) wing intensity,
(c) Ca~II line centre intensity (in logarithmic scale for better
visualisation), (d)~Ca~II Doppler map with a velocity range from
$-2.2$ km s$^{-1}$ (black, towards the observer) to 4.5 km s$^{-1}$
(white, away from the observer), and (e)~Fe~I 617.3~nm LOS magnetogram
in arbitrary units, showing opposite-polarity patches around the pore.
}
\end{figure*}

Horizontal motions of the intensity and Doppler structures in the
working FOV were measured by means of the method of local correlation
tracking \citep[LCT,][]{novsim:1988}. Applied to a time-series of images,
LCT provides a time-averaged horizontal velocity field of all the
structures. We must keep in mind that generally, such a velocity field
does not always reflect the real flow of gas, for example, in a penumbra
\citep{sobotka:1999}. On the other hand, flow characteristics of
photospheric granulation can be retrieved quite well by LCT
\citep{Verma:2013}. We used the filtered data, the {\it FWHM} of
the Gaussian tracking window was set to 1\farcs 2, and the temporal
integration was made over an interval of 52 minutes.

\section{Chromospheric filamentary structure}\label{sec:filam}

\subsection{Morphology}\label{subsec:mor}

Examples of filtered intensity maps (WL, blue wing $-60$ pm from
the line centre, and the Ca~II 854.2 nm centre), a Ca~II Doppler
map and a LOS magnetogram are shown in Fig.~\ref{fig:exa}. The
magnetogram was calculated in arbitrary units by integrating the
red lobe of the Fe~I 617.3~nm $V$ profile and the pore's magnetic
signal (black in Fig.~\ref{fig:exa}e) was saturated at 15~\% of
its maximum for better visualisation. Opposite-polarity patches
(white) are clearly visible in the photosphere around the pore.
In the Ca~II wing we see the typical pattern of reverse granulation
together with magnetic bright points. Short bright filaments surround
the pore's edge. These filaments are a manifestation of an inclined
magnetic field that extends beyond the pore's visible boundary in
the form of spines \citep{SobMorEtal:2012}.

In the line centre, the pore is surrounded by a filamentary structure
composed of radially oriented bright and dark fibrils. This
structure can be detected in the spectral range $\pm 18$~pm around
the line centre. A similar structure is also seen in Doppler maps.
The area and shape of the filamentary structure are identical in all
pairs of the line-centre intensity and Doppler images in the time-series.
It is expected that this structure is formed by magnetic field,
extending into the chromosphere, which connects the pore with
the observed photospheric patches of opposite magnetic polarity.

The fibrils begin immediately at the umbral border and in many
cases continue to the border of the filamentary structure. Their
lengths are in the range 5\arcsec--19\arcsec ~in the line centre and
in the Doppler maps. Widths of the fibrils were measured by
means of power spectra calculated from intensity and Doppler velocity
variations along four quasi-circular curves perpendicular to the fibrils
at different distances from the pore. All power spectra are very similar
each to other. They do not show any particular peaks, instead, they
decrease continuously until 0\farcs 5, where they reach the noise level.
This means that the fibrils do not have any typical width detectable
within the limits of our spatial resolution. The narrowest fibrils
measured directly in the best-quality intensity and Doppler images are
0\farcs 5 (3 pixels) wide.

The fibrils seen in the Ca~II line centre are spatially uncorrelated
with those in the Doppler maps. This is similar to the situation in
a superpenumbra of developed sunspots, where flow channels of the
inverse Evershed effect are not identical with superpenumbral filaments
\citep{Tsiro:1996}. Thus, it is worth checking whether the inverse Evershed
effect is also detectable in the filamentary structure around the pore.

\begin{figure}
  \centering
  \includegraphics[width=6.2cm]{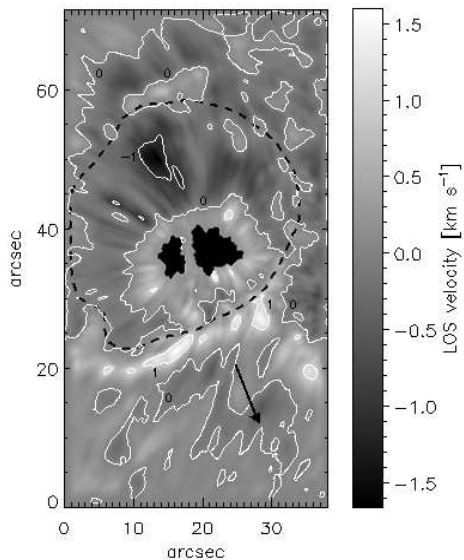}
  \caption[]{\label{fig:Dop}
Time-averaged Doppler map with contours of $-1$, zero, and 1~km~s$^{-1}$.
The black dashed line outlines the border of the chromospheric filamentary
structure around the pore. The arrow is directed towards the disc centre.
}
\end{figure}

\begin{figure*}
  \centering
  \includegraphics[width=18cm]{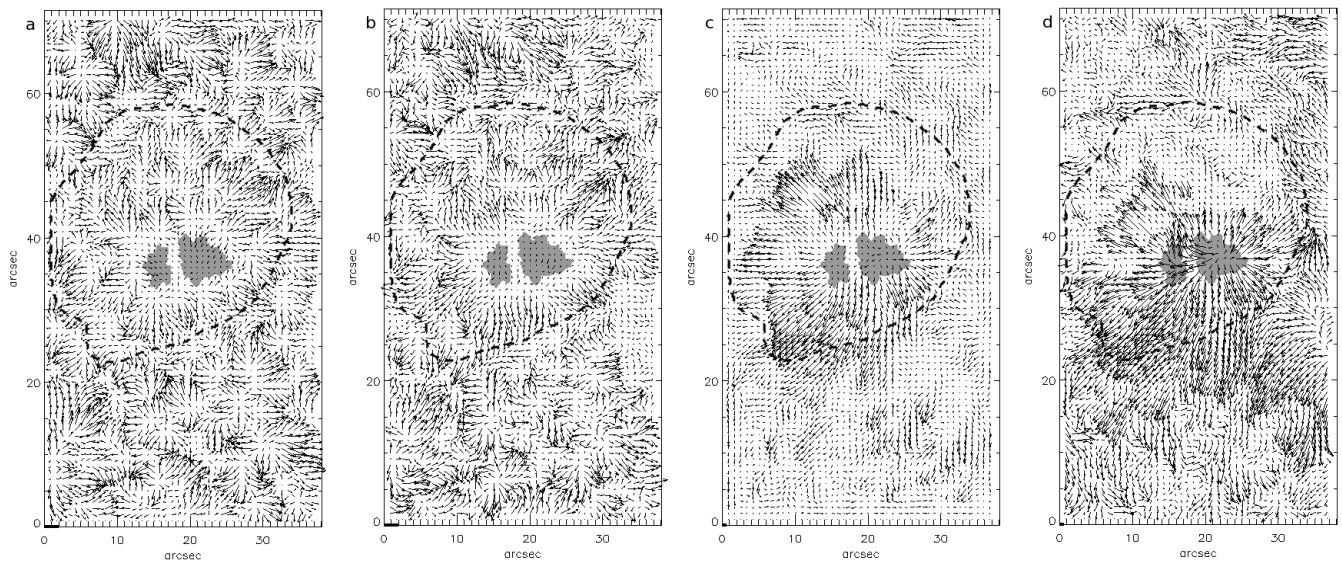}
  \caption[]{\label{fig:LCT}
Horizontal velocities of intensity and Doppler structures, integrated
over a time interval of 52 minutes.
From left to right: (a) broad-band (WL) around 621~nm, (b) Ca~II 854.2 nm
blue ($-60$ pm) wing, (c) Ca~II line centre, and (d) Doppler structures.
The black dashed line outlines the border of the chromospheric filamentary
structure. The black bars at the beginning of the $x$-axis correspond to
1~km~s$^{-1}$ (note that their length is three times longer for (a,b)
than for (c,d)).
}
\end{figure*}

\subsection{Line-of-sight velocities}\label{subsec:LOS}

The statistical distribution of LOS velocities in the $k$--$\omega$
filtered series of Doppler maps, where everything except for the
filamentary structure was masked out, is practically Gaussian,
symmetric around zero, with $\sigma = 0.40$~km~s$^{-1}$. The minimum
and maximum values are $-2.5$ and 4.7~km~s$^{-1}$, respectively.
We adopted the sign convention where positive LOS velocities are directed
away from the observer. The unfiltered LOS velocities, including
oscillations and waves, also show a Gaussian distribution symmetric
around zero, but its $\sigma = 0.65$~km~s$^{-1}$ is larger and the
minimum-to-maximum range is $-3.4$--5.0~km~s$^{-1}$.

The filtered time-series of Doppler maps was averaged in time to
obtain the spatial distribution of mean LOS velocities around the pore.
The result is shown in Fig.~\ref{fig:Dop} together with contours
of $-1$, zero, and 1 km~s$^{-1}$. We can see from the figure that the
inner part of the filamentary structure contains a positive LOS velocity
(away from us, a downflow), while the outer part, located mostly on the
limb side in the north, shows a negative LOS velocity (towards us,
an upflow). Taking into account the heliocentric angle $\vartheta = 23$\degr
~and assuming that the plasma moves along magnetic field lines that
form a funnel, the negative LOS velocity is only partly generated by real
upflows, but mostly by horizontal inflows into the pore. The picture of
plasma moving towards the pore and flowing down in its vicinity is consistent
with the inverse Evershed effect observed in the sunspots' superpenumbra.
This fact, together with the missing correlation between the fibrils
observed in the line-centre and Doppler maps, leads us to the conclusion
that the filamentary structure observed in the chromosphere above the
pore is equivalent to a superpenumbra of a developed sunspot. The
presence of running waves and a rapid change of oscillation frequencies
at the boundary of the pore (see Section~\ref{sec:osc}) further supports
this inference.

\subsection{Horizontal velocities}\label{subsec:horvel}

Apparent horizontal motions of intensity and Doppler structures in the
working FOV were measured using LCT (Section~\ref{sec:proc}) in the
filtered series of images.
The resulting maps of horizontal velocities in WL, Ca~II 854.2 nm blue
($-60$ pm) wing, Ca~II line centre, and of Doppler structures are shown
in Fig.~\ref{fig:LCT}. In the WL map we see a typical \emph{rosetta}
pattern of centres of diverging motions caused by repeated expansion
and splitting of granules and commonly associated with mesogranules
\citep[e.g.][and references therein]{Bonet:2005}. The velocity
magnitudes range from zero to 1.4 km~s$^{-1}$ with a mean value of
0.4 km~s$^{-1}$. Divergence centres just around the pore produce weak
inflows of granules across the pore's border, as was observed by
\citet{sobetal:99}.

In the Ca~II wing (Fig.~\ref{fig:LCT}b) we see a pattern similar to
that in WL, which, in this case, describes horizontal motions in the
reverse granulation.
To our knowledge, such motions were never measured before.
The velocity magnitudes are equal to and the divergence centres are
approximately co-spatial with those in the WL granulation. The most
important difference is that the inflows into the pore are absent.
This can be explained by the fact that  close to the pore's border,
the pattern of reverse granulation is replaced by short bright
filaments (Section~\ref{subsec:mor}) that do not show any
horizontal motions.

The typical apparent motion of chromospheric structures seen in the
Ca~II line centre and Doppler maps is directed radially away
from the pore. The outflow area in the line centre is smaller than the
area of the filamentary structure (Fig.~\ref{fig:LCT}c) and the
spatial distribution of velocity magnitudes is irregular, with a maximum
of 5--6~km~s$^{-1}$ (limited by the cutoff of the $k$--$\omega$ filter)
south-west from the pore. The direction of the outflow is
opposite to the real flow of gas, the inverse Evershed effect, derived
from the LOS velocities. A visual inspection of movies composed of
line-centre images reveals elongated diffuse bright and dark patches
with sizes of 1\arcsec--3\arcsec ~moving intermittently along the
chromospheric fibrils away from the pore.

Motions of the Doppler structures (Fig.~\ref{fig:LCT}d) diverge from
the centre of the large umbral core until 3\arcsec--4\arcsec ~from
the pore's border with a typical speed of 3~km~s$^{-1}$. This type
of motion is probably a residual of running waves
(Section~\ref{subsec:waves}),
the major part of which was removed by the $k$--$\omega$ filter.
A strong outflow with an average speed 3~km~s$^{-1}$ and a maximum of
5--6~km~s$^{-1}$, observed in the region south of the pore, may be
related to outward-moving patches with positive and negative LOS
velocities -- some of them are seen in Fig.~\ref{fig:exa}d.
A similar type of flocculent flows in the line H$\alpha$ was described
by \citet{Vissers:2012}.

Generally, the origin of the apparent horizontal outward motion of
line-centre and Doppler structures is unclear. It might be caused
by a combination of moving intensity or Doppler patches and residuals
of running waves, but also by a spurious LCT correlation of chaotic
transversal motions of fibrils.

\section{Oscillations and waves}\label{sec:osc}

\subsection{Observed oscillation properties}\label{subsec:pow}

\begin{figure}
\centering
\includegraphics[width=0.48\textwidth]{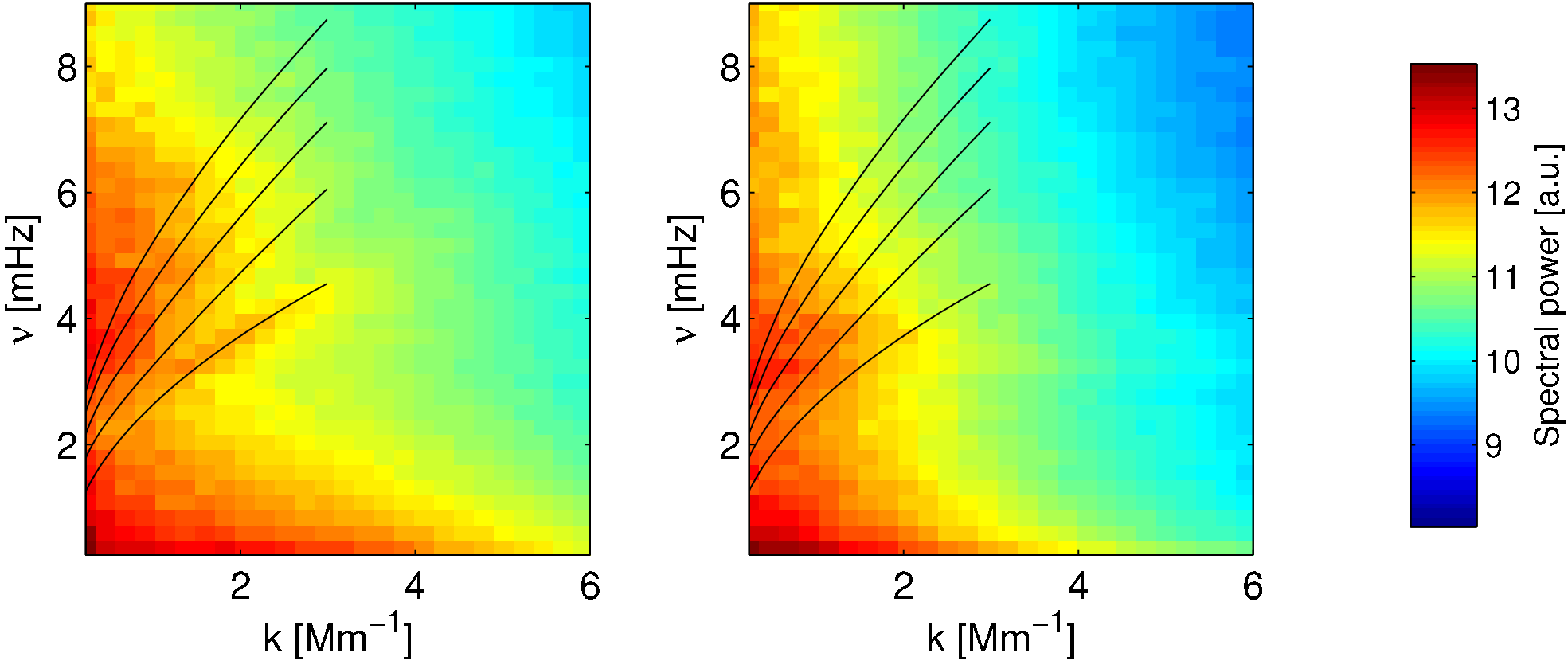}
\caption[]{\label{fig:QS_k-omega}
Comparison of $k$--$\omega$ diagrams obtained for the quiet-Sun
region, constructed for the wings of the Ca~II line (left)
and the centre of the line (right). Model S eigenfrequencies are
overplotted with solid lines.
}
\end{figure}

Before we began to analyse the oscillations in various magnetised
regions, we performed a sanity check of the observations
in the quiet-Sun regions. The spatial extent of quiet-Sun regions
is large enough to construct standard helioseismic
$k$--$\omega$ diagrams. We present them in Fig.~\ref{fig:QS_k-omega},
where we also overplot the eigenfrequencies of the first five ridges
($f$, $p_1$--$p_4$) from Model~S \citep{jcd1996} for reference.
The left diagram, corresponding to the photosphere, is qualitatively
similar to a $k$--$\omega$ diagram constructed from Helioseismic
and Magnetic Imager \citep[HMI;][]{Scherrer:2012,Schou:2012} data,
where we cropped the data cube in time and space to mimic the IBIS
observations.

One can see from the figure that in the wings of the Ca~II line
($\pm 54$ pm from the line centre), where we analyse the signal of
oscillations from the middle photosphere ($h \simeq 200$--300 km),
individual ridges of solar oscillations can be identified, with the
$f$ mode clearly visible and some visibility of $p_1$ and $p_2$ modes.
Higher-order $p$ modes blend into a bulk in the $k$--$\omega$ diagram.
An analogous diagram constructed for the centre of the line, thus
representing the spatio-temporal behaviour of the intensity in the
middle chromosphere ($h \simeq 1300$ km), is very different.
The visibility of the acoustic ridges in the $k$--$\omega$ diagram
drops rapidly because $p$ modes do not propagate through the
temperature minimum in the non-magnetised atmosphere. However,
the power at those frequencies is not zero. This is probably
a consequence of magneto-acoustic waves propagating through
\emph{magnetic portals} into the chromosphere. \citet{Stangal:2011}
reported that magnetic fields are able to lower the cutoff frequency
for acoustic waves, thus allowing the propagation of waves that would
otherwise be trapped below the photosphere into the upper atmosphere.
The power distribution averaged over the wave-number space resembles
that displayed in Fig.~5 of \citet{jefferies2006}.
It is also interesting to look at the spatial power distribution
of plasma motions at frequencies lower than 2~mHz. In the middle
chromosphere, the smaller scales have much less power, in contrast
to the larger scales. This is not surprising, because the dynamics
in the chromosphere must be completely different due to the lower
density and dominance of the magnetic field.

The magnetised regions (the two umbral cores of the pore and the LB) have
a very small spatial extent and an irregular shape, thus we do not discuss
the corresponding $k$--$\omega$ diagram and focus on the properties
of the power spectra in the frequency domain.
The seismic properties of the pore are as expected from the literature
and the eastern and western umbral cores have similar seismic properties.

\begin{figure*}
\sidecaption
\includegraphics[height=6.8cm]{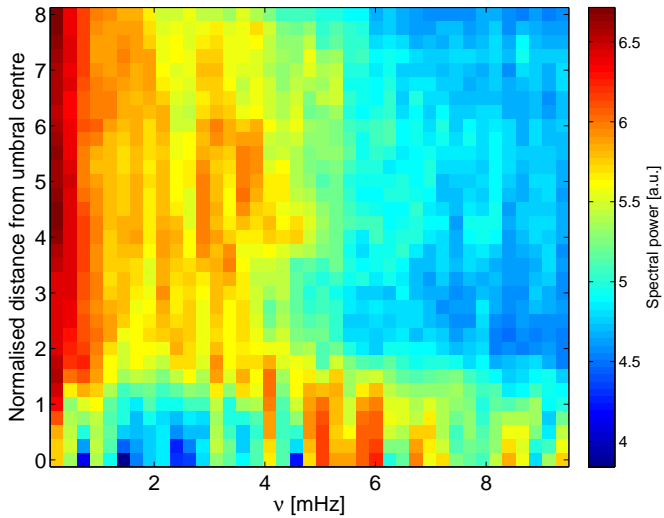}
\rule{3mm}{0pt}
\includegraphics[height=6.8cm]{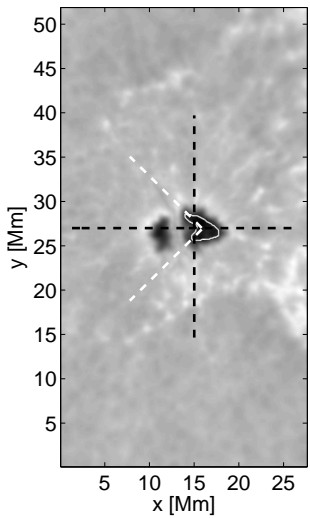}
\caption[]{\label{fig:PS_vs_distance}
Azimuthally averaged power spectra of the Ca~II line-centre
intensity as a function of the normalised radial distance from
the pore's centre and frequency
(left) and a sketch of geometry used (right). The solid line encircles
the \emph{safe boundary} of the pore (see text). The radial distance is
normalised with respect to this boundary (here $R \equiv 1$ for all
angles). Two white dashed lines limit the angles used in the azimuthal
average of the power spectra. The black dashed lines mark the cuts in $x$
and $y$ used to construct time-slice diagrams in Fig.~\ref{fig:slices}.
}
\end{figure*}

For a detailed study of the chromospheric power spectrum we selected
the line-centre intensity in the larger eastern umbral core and
analysed the change of the frequency power-spectrum as a function of
the radial distance. The radial distance $R$ was measured from the gravity
centre of the umbral core ($R=0$) and normalised at each azimuthal angle
$\phi$ to the \emph{safe-boundary} distance $R=1$. This \emph{safe boundary}
was determined by hand from a time-averaged image in the Ca~II wing
(Fig.~\ref{fig:PS_vs_distance} right) and encircles the inner parts of
the umbral core, excluding disturbances from the surroundings. We took 
into account only the directions pointing away from the LB, as shown
in the figure. Then we averaged the frequency power spectra
$P_{\rm I,centre}(\nu,R,\phi)$ azimuthally over all points with
the same normalised distance from the centre of the umbral core.

The result, plotted in Fig.~\ref{fig:PS_vs_distance} left, depicts
dominant frequencies above 4 mHz in the umbra ($R < 1.5$),
while outside the umbra ($R > 1.5$),
all significant frequencies are below 5~mHz. There seems to be
a systematic shift of the peak of the dominant frequency band for
$R < 1.5$, with a power shifting towards lower frequencies when $R$
increases -- in agreement with the findings of \citet{reznikova2011},
suggesting the shift of the acoustic cut-off frequency with increasing
inclination of the field lines. At $R \sim 1.5$ (the visible umbral
boundary), the frequencies higher than 4~mHz essentially disappear from
the power spectrum and all the power is shifted towards 2--3 mHz with a
second power bulk at frequencies below 1~mHz, where plasma motions dominate.
This is similar to the change in frequencies found by \citet{tziotzio2007}
at the penumbra-superpenumbra boundary (Fig.~6 of that paper). Therefore,
the filamentary region around the pore depicts a seismic behaviour in the
chromosphere similar to that of the superpenumbra of a sunspot.

\begin{figure}
\centering
\includegraphics[width=7cm]{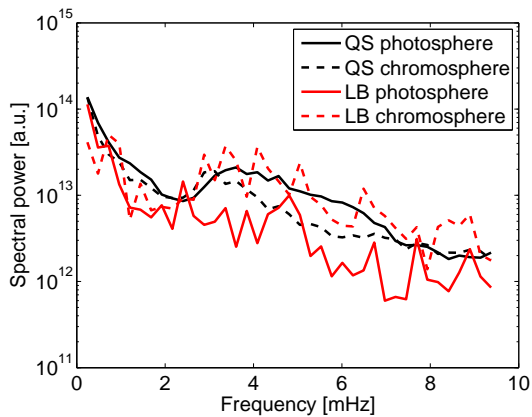}
\caption{\label{fig:PS_LB_QS}
Average power spectra of the quiet-Sun (QS) regions in the FOV
constructed from the Ca~II wings (photosphere) and the line centre
(chromosphere), compared with the power spectra of the light bridge (LB)
in the photosphere and chromosphere.
}
\end{figure}

It is interesting, however, to look at the seismic properties of the LB
and compare them with the quiet Sun, because the magnetic-field strength
is significantly lower in the LB than in the surrounding umbral cores
of the pore. The power spectra averaged over all pixels belonging to
the LB and quiet-Sun regions are displayed in Fig.~\ref{fig:PS_LB_QS}.
The power spectra of the LB are noisier because there forty times fewer
pixels enter the averaging than in the quiet Sun. Interestingly,
the power spectrum of the LB observed in the chromosphere is very
similar to the photospheric power spectrum of the quiet Sun with
a small drop of power below 2~mHz. To explain this observation, we
suggest that the $p$ modes, which are normally evanescent in the
chromosphere, just leak into the chromosphere along the inclined
magnetic field in the LB, which was also discussed by other authors
\citep[e.g.,][]{Cally:2006,depontieu2004}. We assume that the lower
frequency power is missing because although the LB has some properties
similar to the solar photosphere, it is not the photosphere,
and especially the convection is still operating in a degenerated
state, which affects the conditions for the wave excitation.

\begin{figure*}
\centering
\includegraphics[width=0.92\textwidth]{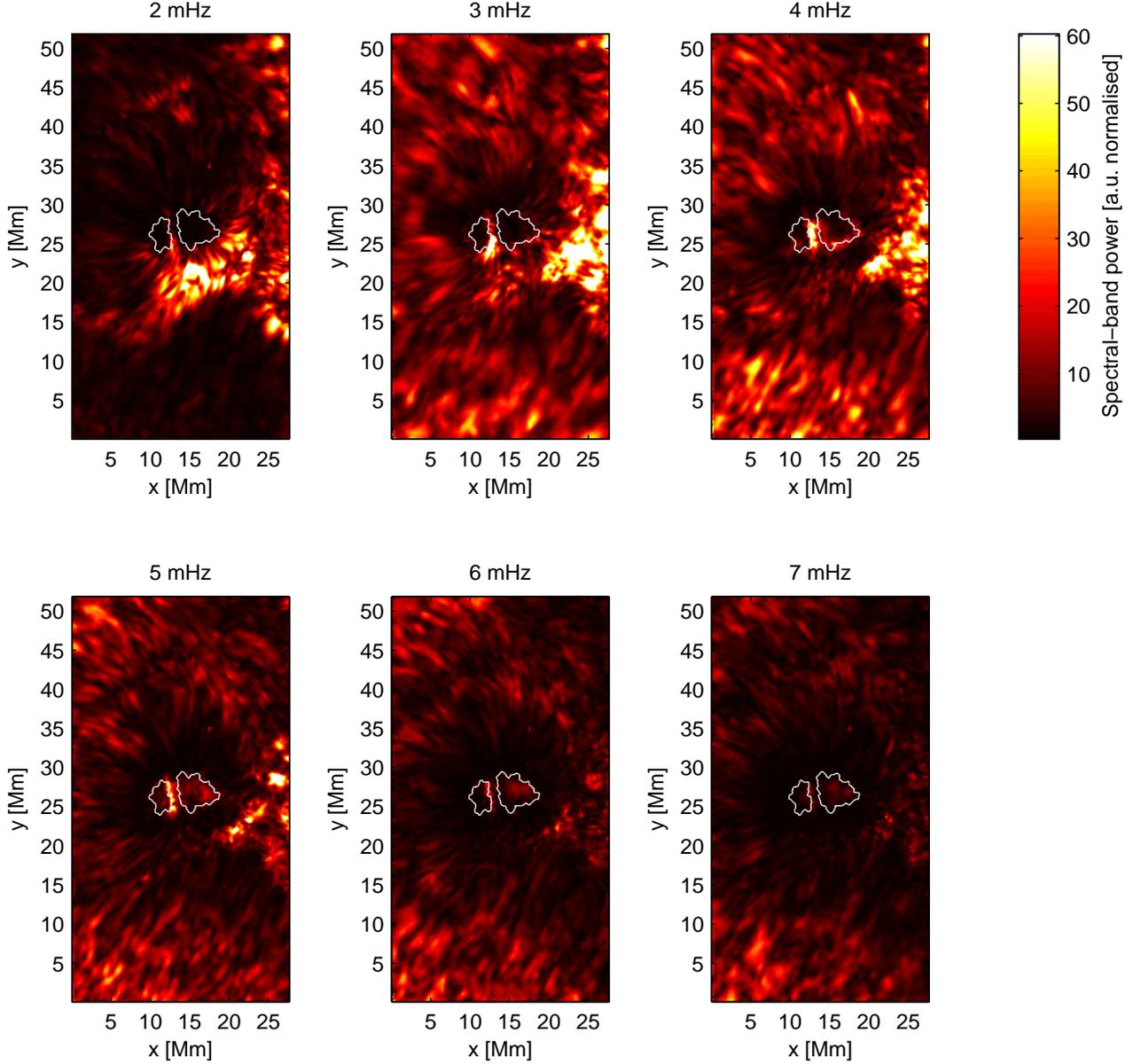}
\caption{\label{fig:powermaps}
Acoustic power maps measured from the series of Dopplergrams in six
frequency bands around 2--7~mHz with {\it FWHM} of the transmission
band 0.86~mHz. All panels have the same colour scale.
}
\end{figure*}

From the time-series of Dopplergrams we constructed power maps
for nine frequency bands around 1, 2, ..., 9~mHz with a transmission
{\it FWHM} = 0.86~mHz. Six of these are displayed in Fig.~\ref{fig:powermaps}.
These power maps nicely show the changing spectral distribution of the
acoustic emission. A region south of the pore is dominant in the
low-frequency bands, which is probably an artefact caused by horizontal
motions of Doppler structures (cf. Fig.~\ref{fig:LCT}d). The spectral
power in frequency bands around 3 to 9~mHz is significantly lower in
the filamentary region around the pore than in the surroundings.
The acoustic emission in both umbral cores peaks around 5--7~mHz,
that is, the band of three-minute oscillations. This agrees with
\citet{Stangal:2012}, who showed that the three-minute enhancement is
strictly confined to the umbral region, while five-minute waves are
suppressed here. Around 4~mHz, the power is dominant
in the LB, a plage eastward of the pore, and at the edges of umbral
cores. We assume that these are the regions where $p$ modes leak up
along the inclined magnetic field into the chromosphere.

\subsection{Umbral flashes and running waves}\label{subsec:waves}

\begin{figure}
  \centering
  \includegraphics[width=8cm]{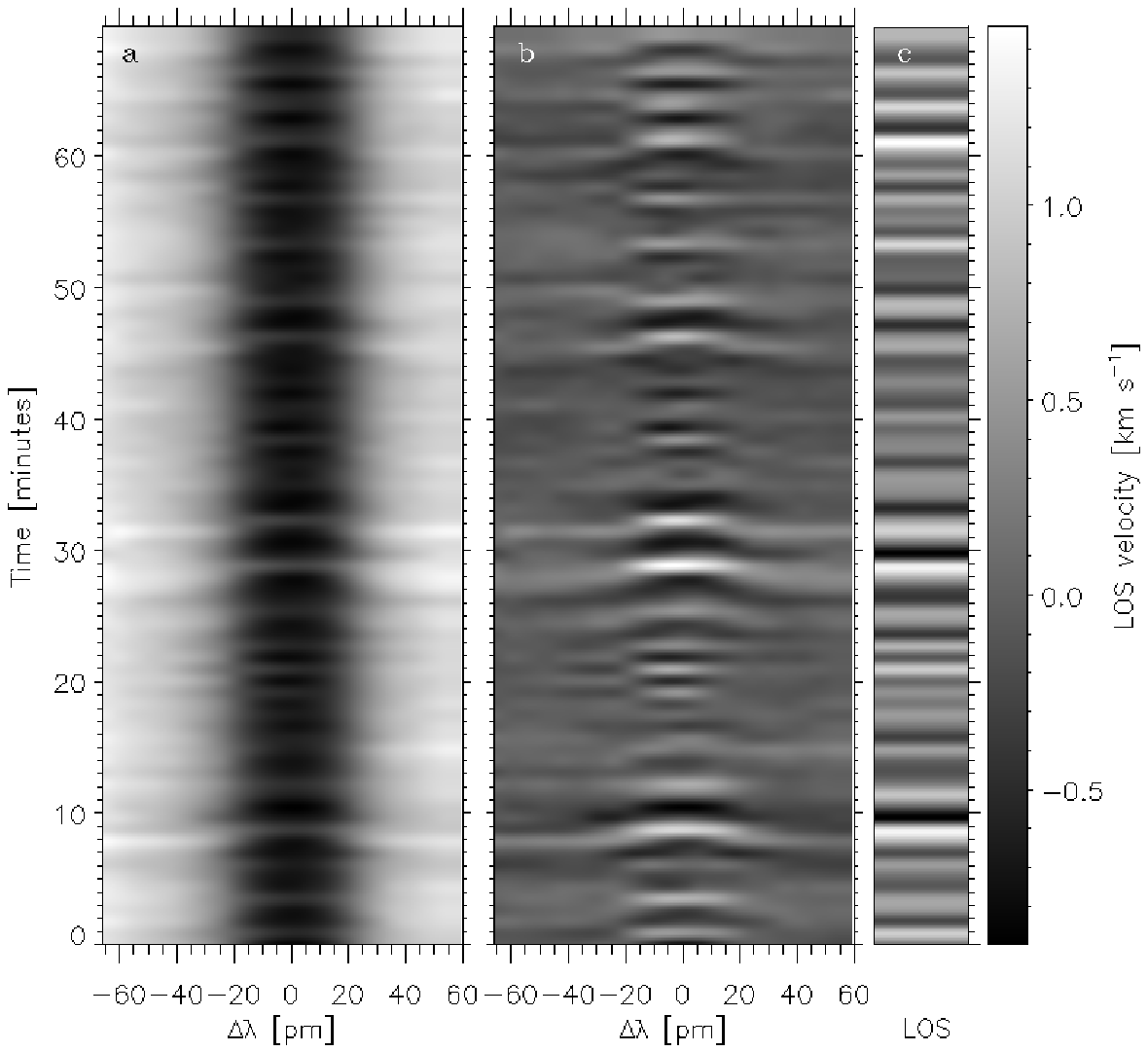}
  \caption[]{\label{fig:umfl}
Oscillations in the centre of the pore's large umbral core. Prominent
umbral flashes occur at $t = 9$, 29, and 32 minutes.
(a) Temporal fluctuations of intensity at different wavelengths in
the Ca~II line profile.
(b) Same as (a) but enhanced by subtraction of the mean line profile.
(c) Visualised temporal fluctuations of the Doppler (LOS) velocity.
Redshifts are bright.
}
\end{figure}

Umbral flashes are seen in the Ca~II 854.2 line core. They appear in
the two umbral cores of the pore and typically cover the whole area
of each core. For a more detailed analysis we selected a small
0\farcs 5 $\times$ 0\farcs 5 area in the centre of the dominant eastern
umbral core and constructed a time-vs.-wavelength diagram of temporal
fluctuations of intensity in the observed Ca~II line profile. This
diagram, together with fluctuations of the Doppler velocity visualised
in a grey scale, is shown in Fig.~\ref{fig:umfl}. We can see that
umbral flashes appear as intensity enhancements of chromospheric umbral
oscillations and are connected with an abrupt change from a redshift to
a blueshift in the Doppler signal \citep{Rouppe:2003}.

Photospheric counterparts of umbral flashes are also seen in the
Ca~II line wings. From the cross-correlation analysis we obtain
that intensity brightenings in the $\pm 54$ pm wings typically
precede umbral flashes by 56 s. Assuming that umbral flashes
are excited by upward-propagating shock waves \citep{Rouppe:2003}
and that the height difference between the formation levels of
the Ca~II line wings and centre is 1000--1200~km, we obtain a
mean phase-speed of the upward propagation of approximately
20~km~s$^{-1}$. This speed is supersonic. The sound speed
\begin{equation}
c_{\rm s} = \sqrt{\gamma p/\rho},
\end{equation}
where $\gamma$ is the adiabatic index equal to 5/3 for monoatomic gas,
$p$ is the gas pressure, and $\rho$ is the gas density,
was estimated using the quiet-Sun model {\it VAL3C} \citep{VAL3:1981}
and the umbral model {\it M} \citep{Maltby:1986}. In the middle
photosphere, $c_{\rm s} \simeq 7$ and 6~km~s$^{-1}$ for the quiet Sun
and umbra, respectively, and in the middle chromosphere,
$c_{\rm s} \simeq 8$ and 9~km~s$^{-1}$.

Concentric running waves originating in the centre of the dominant
eastern umbral core and propagating beyond the pore's edge into the
chromospheric filamentary structure are observed in the unfiltered
time-series of Ca~II intensity and Doppler maps. In the intensity,
the waves start to be detected in the line wings for
$|\Delta \lambda| < 30$~pm, i.e., in the part of the line profile formed
in the chromosphere. The waves are best observable in the line centre.
The shapes of waves observed visually in pairs of line-centre intensity
images and Dopplergrams are practically equal.

\begin{figure}
  \centering
  \includegraphics[width=8.5cm]{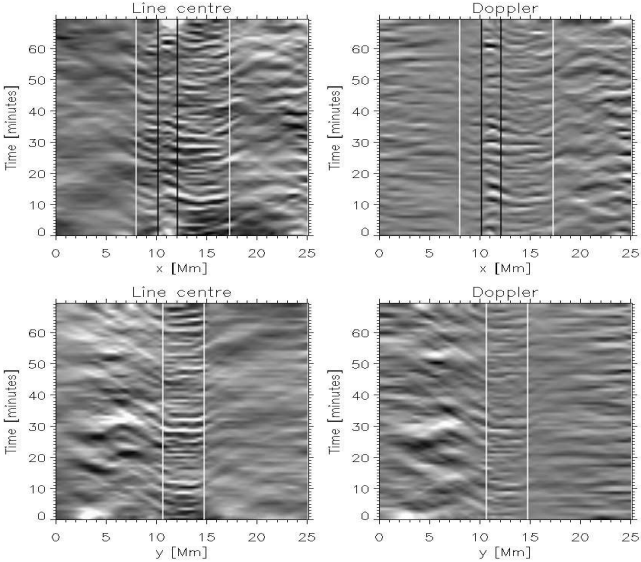}
  \caption[]{\label{fig:slices}
Time-slices showing umbral flashes and running waves observed in
difference maps (unfiltered $-$ filtered) of the
Ca~II line-centre intensity (left) and Doppler velocity (right). The
slices are constructed along the cuts in $x$ (top) and $y$ (bottom)
directions displayed in Fig.~\ref{fig:PS_vs_distance}. White lines
mark edges of the umbra, black lines the borders of the light bridge.
The grey scale of Doppler velocities ranges from $-2$ to 2~km~s$^{-1}$.
}
\end{figure}

To measure their speed and range of propagation, we used two time-series
of difference maps of line-centre intensities and Doppler velocities,
obtained by subtraction of the $k$--$\omega$ filtered images from the
original ones. Then we constructed time-slice diagrams from two
perpendicular cuts intersecting in the centre of the eastern umbral core
(see Fig.~\ref{fig:PS_vs_distance}, right). The time-slice diagrams
are displayed in Fig.~\ref{fig:slices}. Note the waves propagating
through the smaller western umbral core (top left panel)
and the strong Doppler velocity response in the LB with maximum amplitude
$\pm 4$~km~s$^{-1}$ to umbral flashes (Fig.~\ref{fig:slices}, top right).
The mean phase velocity of running waves measured in the Ca~II line-centre
intensity is 9~km~s$^{-1}$, comparable to the sound speed in the
chromosphere. The mean phase velocity derived from Dopplergrams is
10~km~s$^{-1}$. The waves pass the pore's edge and propagate through
the chromospheric filamentary structure to the distance of typically
3~Mm from the visible border of the pore, in extreme cases to 5--6~Mm,
and then disappear. These waves are very similar to the running penumbral
waves observed in the penumbral chromosphere and superpenumbrae of
developed sunspots, described for instance by \citet{Christop:2000}.

\section{Light bridge}\label{sec:LB}

\subsection{Structure and width}\label{subsec:LBstruct}

The strong granular LB separates the two umbral cores of the pore.
It is the brightest feature inside the pore. In the chromosphere,
it is by factor of 1.3 brighter than the average brightness in the
working FOV. The granular structure of the LB is preserved at all
heights, from the photospheric continuum level to the formation
height of the Ca~II line centre. It is interesting that while a
typical pattern of reverse granulation, observed in the Ca~II wings,
appears outside the pore in the middle photosphere, the LB is always
composed of small bright granules separated by dark intergranular
lanes (see Fig.~\ref{fig:exa}). This fact is discussed in
Section~\ref{sec:disc}.

A feature-tracking technique \citep{sbs:1997} was applied to correlate
the LB granules in position and time at different heights in the
atmosphere. A correlation was found between the photospheric LB granules
in the WL and Ca~II wing ($\pm 60$~pm) images with a correlation
coefficient of 0.46. On the other hand, there was no correlation
between the chromospheric LB granules observed in the Ca~II line
centre and the photospheric ones in the wings and continuum.
The feature-tracking technique also showed that the mean size of the
LB granules increased with height from 0\farcs 45 in the WL images to
0\farcs 50 in the Ca~II wings and 0\farcs 54 in the Ca~II line centre.
Similarly, the average width of the LB increased with height in the
atmosphere from 2\arcsec ~(WL) to 2\farcs 5 (Ca~II wings) and 3\arcsec
~(Ca~II line centre). This contradicts the expectation that the
width of the LB magnetic structure decreases with height due to the
presence of magnetic canopy.

\begin{figure*}
\centering
\includegraphics[height=7.5cm]{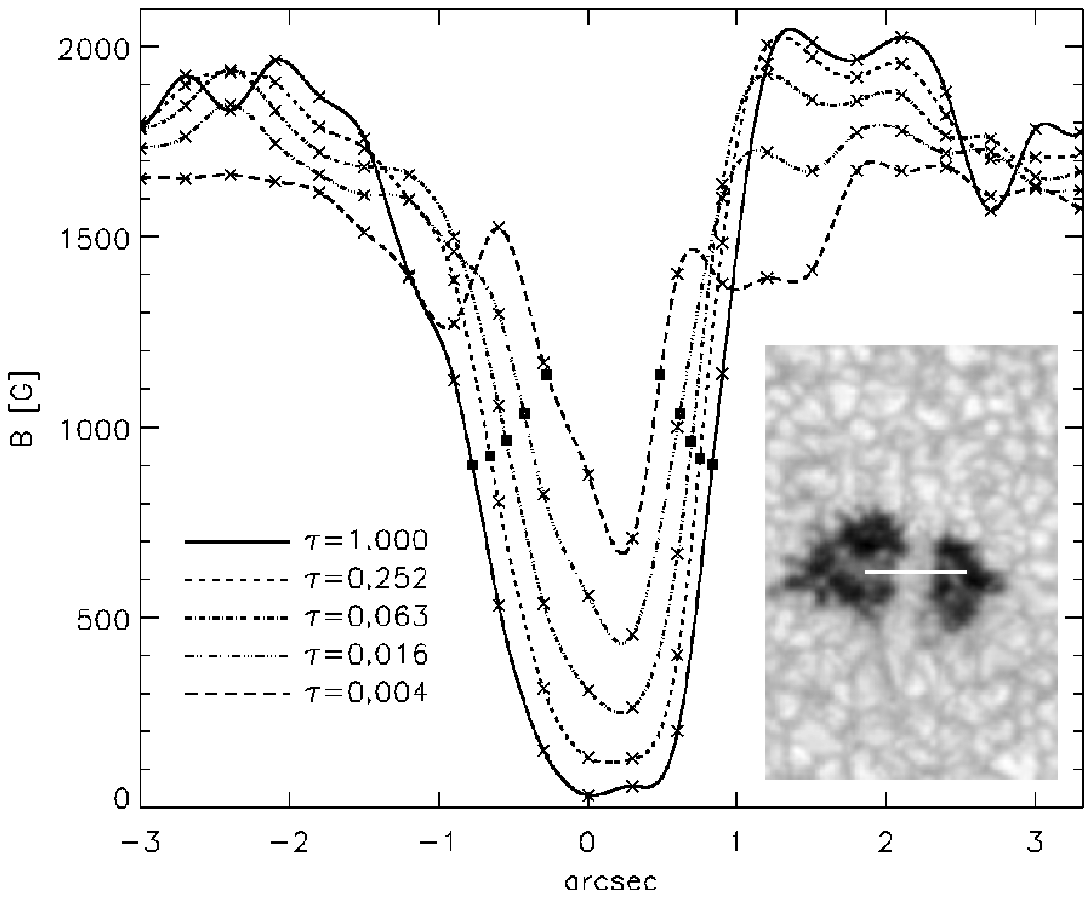}
\includegraphics[height=7.5cm]{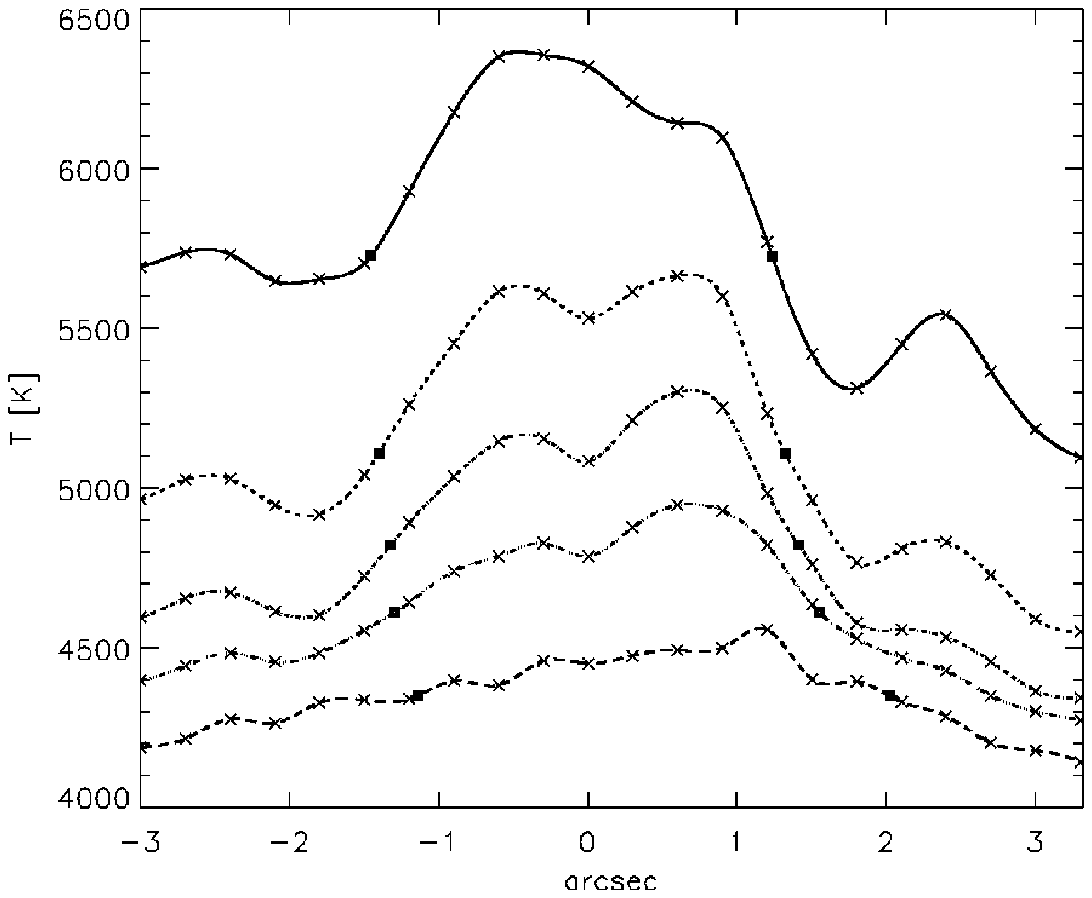}
\caption[]{\label{fig:LBwidth}
Profiles of magnetic-field strength $B$ (left) and temperature $T$
(right) across the light bridge at different optical depths $\tau_{500}$,
obtained by inverting HINODE observations. The location of the cut is
shown in the inset. Crosses denote data points obtained by the inversion
code and black squares the positions used to measure the width.
}
\end{figure*}

To investigate the width of the LB at different heights in
the photosphere, we used the complementary HINODE observations in the
two spectral lines Fe~I 630.15 and 630.25 nm, made 3 hours 14 minutes
before our observations started, and applied the inversion code SIR
(see Section~\ref{sec:proc}) to retrieve vertical stratifications of
temperature and magnetic-field strength from the observed Stokes
profiles. From the resulting stratifications we created maps at
five different optical depths from $\tau_{500} = 0.004$ to 1
with steps of $\log (\tau_{500}) = 0.6$. We cut across the central
part of the LB and measured the width of the structure in maps
of temperature (corresponding to the intensity at each height) and
magnetic-field strength. The width was determined as a {\it FWHM}
related to the highest and lowest values in the cut.
Since the HINODE pixel scale 0\farcs 3 is insufficient to study
the changes of LB width, we used the spline function to refine
the spatial sampling ten times.
In Fig.~\ref{fig:LBwidth} we present changes of magnetic field
strength and temperature along the cut through the pore and the LB.
Different curves correspond to five different optical depths. Dips
in the temperature profile of the LB at optical depths
$\tau_{500} = 0.016$--0.252 correspond to a central dark lane, seen
also in the wings and centre of the Ca~II line (Fig.~\ref{fig:exa}).
The measured widths are summarised in Table~\ref{tab:1} together with
geometrical heights and lowest values of the magnetic-field strength
in the LB, obtained from the inversions.

\begin{table}
\caption{
Widths of the light bridge measured from temperature and magnetic field
maps (HINODE observations) at different optical depths $\tau_{500}$ in
the photosphere; $h$ is the geometrical height and $B_{\rm min}$ the
lowest magnetic-field strength in the light bridge.}
\label{tab:1}
\centering
\begin{tabular}{rrrrr}
\hline\hline
\noalign{\smallskip}
$\tau_{500}$ & $h$ [km] & $T$ width & $B$ width & $B_{\rm min}$ [G]\\
\hline
\noalign{\smallskip}
1.000 &   0 & 2\farcs 7 & 1\farcs 6 & 50\\
0.252 &  90 & 2\farcs 7 & 1\farcs 4 & 150\\
0.063 & 190 & 2\farcs 7 & 1\farcs 2 & 250\\
0.016 & 280 & 2\farcs 9 & 1\farcs 1 & 450\\
0.004 & 360 & 3\farcs 2 & 0\farcs 8 & 700\\
\hline
\end{tabular}
\end{table}

The width of the LB in temperature in the low photosphere is approximately
constant (2\farcs 7). At higher layers, the LB becomes wider. However,
at the highest studied optical depth, the temperature contrast between the
pore and the LB is low, which results in a higher error of the measured
width. The temperature response functions are also close to zero at this
optical depth, that is, the retrieved temperature values have higher
uncertainties than deeper photospheric layers.
On the other hand, the light bridge is becoming narrower with height in
the maps of the magnetic-field strength $B$. This corresponds to the magnetic
canopy above the light bridge, which was observationally confirmed by
\citet{JurLB:2006}. The magnetic field from the surrounding umbral cores
is expanding above the LB's weaker field, forming a cusp-like structure
at the height $h \simeq 220$ km \citep{giordano:08}. Note that the central
part of the LB is almost field-free at the lowest photospheric layer.
However, the response function to $B$ is close to zero there and the
retrieved values of $B$ have high errors ($\pm 250$ G). At the
highest optical depth, the uncertainties of $B$ are also of this order.
The errors at other optical depths are typically $\pm 100$ G.

As a whole, the LB centre is shifting towards the smaller umbral core
(see Fig.~\ref{fig:LBwidth}). This is caused by the more inclined magnetic
field in the large eastern umbral core, which expands faster above the LB
than the almost vertical magnetic field from the smaller core.
In \citet{SobMorEtal:2012} we determined the magnetic-field inclination
and azimuth in the LB from inversions of Stokes profiles of Fe~I 617.33 nm,
observed simultaneously with the Ca~II line. The azimuth direction is
perpendicular to the LB axis. The magnetic-field inclination is
30\degr--35\degr, measured at $\tau_{500} = 0.03$--0.1, which corresponds
to the height $h$ of about 100 km, that is, below the cusp height.
For the oscillation propagation the inclination above the cusp height
is important, which is composed of the 10\degr ~general tilt of
the pore's field (Section~\ref{sec:obs}) and the inclination of the
dominant expanding field of the large umbral core, extrapolated
to high photosphere ($h > 400$~km). We made this extrapolation using
straight field lines and obtained a total inclination of 40\degr--45\degr.
If the curvature of extrapolated field lines is taken into account,
the inclination can reach 50\degr.

\subsection{Acoustic flux}\label{subsec:LBosc}

From the preceding section it becomes clear that
the apparent width of the LB is larger than the width of the structure with
reduced magnetic field within the LB. At the same time, the LB in the
chromosphere is by factor of 1.3 brighter than the average brightness in
the FOV. Therefore, there must be some heating mechanism to explain these
observed properties. Oscillations are detected both in the photosphere and
chromosphere, so that we estimated the amount of energy transferred by
oscillations from the photosphere to the higher atmosphere.

Following \citet{bg2009}, we estimated from the Doppler velocities the
acoustic energy flux in the chromosphere of the LB and compared it with
the flux in the quiet chromosphere. The method of power spectra calculation
and calibration in absolute units is described by \citet{Rieutord:2010}. 
It is evident from the power maps in Fig.~\ref{fig:powermaps} that
in the region of the LB, the oscillatory power at frequencies different
from both umbra and quiet chromosphere dominates, while it is similar in
behaviour to the plage region eastward of the pore.
The acoustic power flux can be estimated from the equation
\begin{equation}
F_{\rm ac,tot}=\int\limits_{\nu_{\rm ac}}^\infty \rho P_{\rm v}(\nu)
\upsilon_{\rm gr}(\nu)/TF(\nu)\, {\rm d}\nu,
\label{eq:acousticflux}
\end{equation}
where $P_{\rm v}$ is the velocity spectral power density, $\upsilon_{\rm gr}$
is the group velocity with which the energy is transported, given by
\begin{equation}
\upsilon_{\rm gr}=c_{\rm s}\sqrt{1-(\nu_{\rm ac}/\nu)^2},
\end{equation}
$c_{\rm s}$ is the sound speed from Equation~(1) and $\nu_{\rm ac}$
is the acoustic cutoff frequency given by
\begin{equation}
\nu_{\rm ac}=\frac{\gamma g \cos{\theta}}{4\pi c_{\rm s}},
\end{equation}
with $g$ being the surface gravity and $\theta$ the magnetic field inclination.
The effect of magnetic-field inclination that lowers the acoustic cutoff
frequency was analytically derived by \citet{Cally:2006}.

For the model atmosphere used to derive the acoustic cutoff and group velocity
we adopted the values of $\rho=6.5\times 10^{-8}$~kg\,m$^{-3}$ and $p=2.5$~Pa
from the {\it VAL3C} model \citep{VAL3:1981} averaged over heights of
900--1200~km. We expect that this model an of average quiet chromosphere
also provides reasonable values for the LB.
We estimated the field inclination $\theta=50$\degr ~by extrapolating
the inversion results (Section~\ref{subsec:LBstruct}) obtained for the
photospheric layers.

\begin{figure}
\centering
\includegraphics[height=7.5cm]{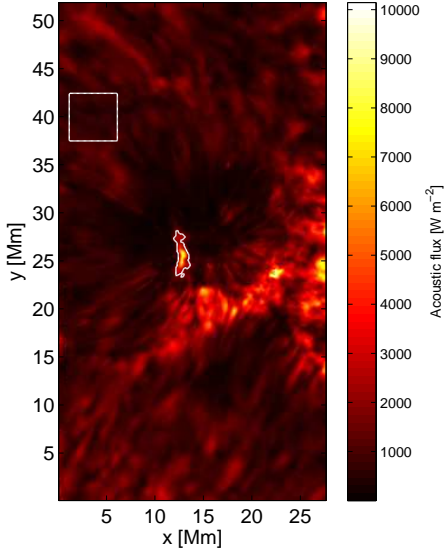}
\caption[]{\label{fig:powermaps_selection}
Map of the total acoustic power flux with the selection of the
light-bridge region and the region of the quiet chromosphere (square).
}
\end{figure}

\begin{figure}
\centering
\includegraphics[width=7cm]{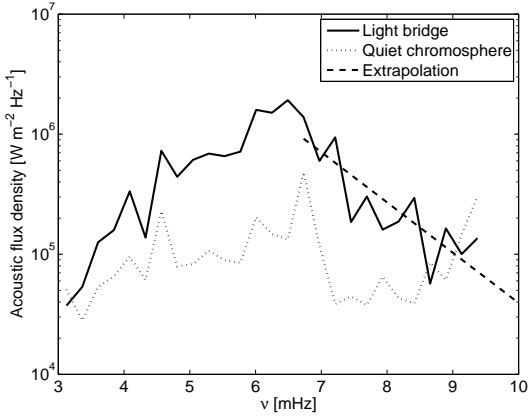}
\caption[]{\label{fig:acoustic_power}
Frequency distribution of the spatially averaged acoustic power flux
in the light bridge and the quiet chromosphere.
}
\end{figure}

The function $TF(\nu)$ in the equation given above denotes the
\emph{transfer function} of the solar atmosphere. The extent in
atmospheric heights of the spectral-line contribution function decreases
the observable signal of small-scale fluctuations along the line of sight,
thus the signal with shorter periods is usually retrieved by observations
with decreased power. For the estimate in this paper we did not attempt to
derive the correct $TF$ and set $TF(\nu)=1$ for all frequencies.
The correction using the transfer function may increase the total estimated
power by a factor of two or similar, which, given the other assumptions we
made, is not critical in our estimate.

The spectral power density $P_{\rm v}(\nu)$ was derived from the series of
Dopplergrams computed in the inner wings of the spectral line with the
assumed height in the atmosphere around 1000~km. In the map we then
segmented out the region of the light bridge and also the region of the
quiet chromosphere for comparison (Fig.~\ref{fig:powermaps_selection}).
The total acoustic power flux averaged over the LB region measured using
our assumptions is 3400~W\,m$^{-2}$, while only 680~W\,m$^{-2}$ in the
quiet-chromosphere region. One can easily note that the peak acoustic
flux within the LB reaches values as high as 10\,000~W\,m$^{-2}$ averaged
over  an observation pixel. These peak values are much higher than the
value of 1840~W\,m$^{-2}$ reported by \citet{bg2009}, obtained for the
quiet photosphere for uncorrected data, that is, with $TF(\nu)=1$.
It seems that in the quiet-Sun regions, some of the acoustic flux of
the $p$ modes is transferred to the chromosphere, possibly along the
inclined magnetic field at supergranular boundaries. However, in
the studied LB and similarly in the plage region
(see Fig.~\ref{fig:powermaps_selection})
the $p$-mode power flux into the active chromosphere is much higher
and probably contributes to the plasma heating in these regions.

Again following \citet{bg2009}, we extrapolated the high-frequency tail
of the $P_{\rm v}(\nu)$ using a power law to account for acoustic flux
caused by the high-frequency waves (Fig.~\ref{fig:acoustic_power}).
The correction is very small however. Taking the estimated tail
contribution into account, the total spatially averaged acoustic flux
is 3460~W\,m$^{-2}$, thus the extrapolated tail adds only a negligible
2~\% of the flux.

We note that the estimate of the acoustic power might be biased because of
the inaccuracies in the modelling. For example, the map of the acoustic
flux in Fig.~\ref{fig:powermaps_selection} was computed assuming the
{\it VAL3C} atmosphere at each pixel, which certainly is inappropriate. We
did not attempt to appropriately model the atmosphere in each pixel of the
map, since we have measurements only in one spectral line. However, to
investigate the sensitivity of the derived acoustic flux to atmospheric
parameters we repeated the calculations using a set of semi-empirical
atmospheric models ({\it VAL3A, VAL3B, VAL3C,} Maltby-{\it M}). The
resulting numbers vary by as much as 30\%, which we estimate to be the
level of our uncertainty. Another uncertainty comes from the unknown
transfer function $TF$ of the solar atmosphere, to derive which,
a detailed  time-dependent model of the atmosphere is needed.
Note that incorporating the correct transfer function increases the
estimate for the acoustic power flux. \citet{bg2009} found a correcting
factor of 1.7 for a photospheric spectral line. Accordingly, our numbers
should be seen as qualitative estimates -- the real acoustic power flux
may be a factor of a few  different from our numbers, probably larger.
Our intention is just to compare the acoustic flux in the light bridge
with the surrounding quiet chromosphere to shed some light on the heat
dissipation process observed in the LB chromosphere.

\subsection{Radiative losses}\label{subsec:LBrad}

To investigate whether the acoustic flux along the inclined magnetic
field above the LB might  be responsible for the heating of the plasma
within the LB chromosphere and consequently for the increased optical
width of the LB, we estimated the radiative losses and compared them with
the estimated heat deposited by the waves. We evaluated the equation
\begin{equation}
\nabla \cdot \left[ F_{\rm ac,tot} + F_{\rm other} \right] = Q,
\label{eq:heatballance}
\end{equation} 
where $F_{\rm ac,tot}$ is the acoustic power flux computed from
Equation~(\ref{eq:acousticflux}), $F_{\rm other}$ is the power flux
from other sources, such as small-scale reconnections \citep{Shimi:2012},
and $Q$ are the radiative losses (net radiative cooling rates). 

\begin{figure}
\centering
\includegraphics[width=8cm]{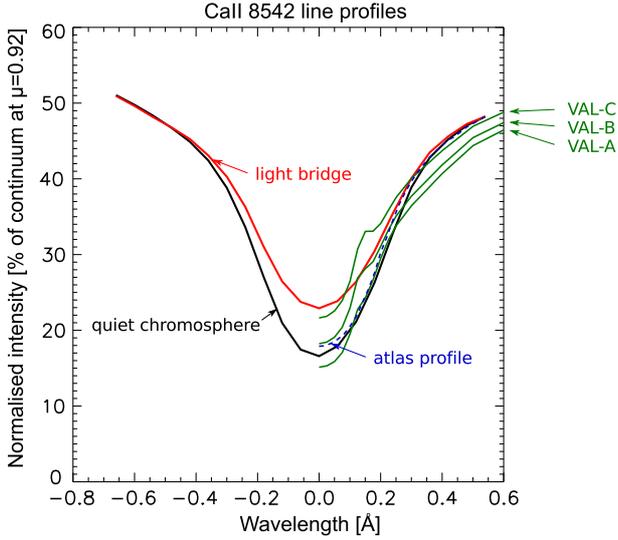}
\caption[]{\label{fig:lineprofiles}
Profiles of the line Ca II 854.2 nm. The observed profiles in the light
bridge (red) and quiet chromosphere (black) are averaged in time and over
the corresponding areas marked in Fig.~\ref{fig:powermaps_selection} and
are normalised to the atlas profile (blue dashes) in the line wing. Green
lines show profiles calculated using the models {\it VAL3A, B,} and {\it C}.
}
\end{figure}

To consider the radiative losses, we made a rough estimate based on the
standard semi-empirical models by \citet{VAL3:1981} that fit our Ca II
profiles quite well. The observed line profiles, averaged in time over
the whole observing period and spatially over the LB and quiet-chromosphere
regions marked in Fig.~\ref{fig:powermaps_selection}, were normalised to
the atlas profile \citep{Linsky:1970} using the far wings of the line,
where both observed profiles coincide. As evident from
Fig.~\ref{fig:lineprofiles}, the spatially averaged profiles may be
reasonably fitted with the {\it VAL3C} model in the LB, and {\it VAL3A}
or {\it VAL3B} models in the quiet chromosphere (from now on we use the
{\it VAL3B} model of the average intranetwork area).
It is well known that the computed Ca~II line profiles are very sensitive
to values of the microturbulent velocity. Higher microturbulence smoothes
the small humps in the near wings, while zero microturbulence leads to a
large enhancement of them; these results are consistent with the Ca~II
modelling by \citet{Uitenbroek:1989}. Since we do not know the realistic
distribution of microturbulent velocities in the LB, which can also be
related to the acoustic-wave dynamics, we just used the standard
{\it VAL3C} distribution. Our computed profile, when convolved with
an instrumental or macroturbulent broadening, can give even a better fit
to the observed LB profile than that shown in Fig.~\ref{fig:lineprofiles}.

\begin{figure}
\centering
\includegraphics[width=8cm]{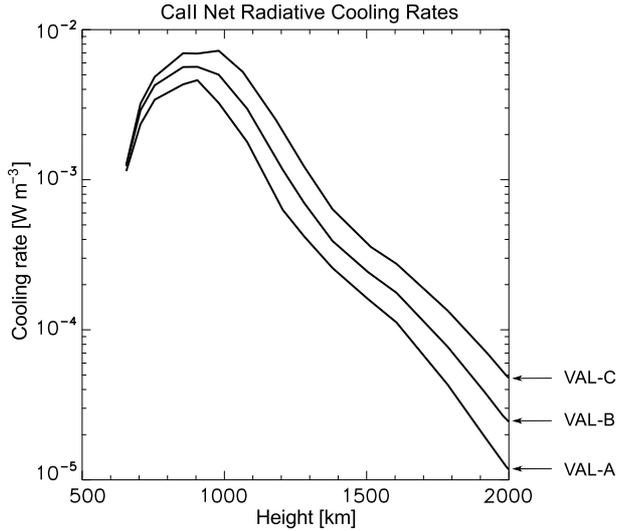}
\caption[]{\label{fig:radiativelosses}
Ca II net radiative cooling rates vs. geometrical height, calculated for
the models {\it VAL3A, B,} and {\it C}.
}
\end{figure}

The Ca~II line profiles and radiative losses were computed for the three
{\it VAL} models,  using a non-LTE atmospheric code based on the MALI
technique \citep{RybHum:1991,RybHum:1992}, with standard partial frequency
redistribution (PRD, angle-averaged) in the CaII~H and K resonance lines.
The hydrogen version of the code (with PRD in Lyman lines) first computes
the ionisation structure and hydrogen level populations, then the Ca~II
version is used with a five-level plus continuum atomic model. In the range
of heights of interest, the hydrogen losses were found to be much smaller than
the Ca~II ones. The Ca~II losses in the H and K lines and in the infrared
triplet, integrated over heights, agree with those computed by
\citet{VAL3:1981} using a similar PRD approach. For the model {\it VAL3C},
these authors also provided integrated losses in the Mg~II h and k lines,
which amount roughly to one third of the Ca~II losses. We use this
estimate in our Table~\ref{tab:energetics}. The Ca~II net radiative
cooling rates are plotted in Fig.~\ref{fig:radiativelosses}
and the total Ca~II radiative losses, that is, the net cooling rates
integrated over all heights displayed in the figure, are given in Table~\ref{tab:radiativelosses}. Note that the dominant contribution
of the integral comes from the heights 700--1400~km; outside of this
interval, the contribution is an order of magnitude smaller.

The acoustic flux was estimated at two heights in the chromosphere:
at $h \sim 1000$~km from the series of Dopplergrams measured in the inner
wings of the line by the double-slit method, and at $h \sim 1400$~km from
the Dopplergrams obtained by the parabolic fit of the line centre.
Hence, by integrating Equation~(\ref{eq:heatballance}) over height $h$ and
neglecting the horizontal variations of the energy flux, as we deal with the
spatially averaged values, we obtain
\begin{eqnarray}
&& \int_{h_1}^{h_2} \frac{{\rm d} \left[F_{\rm ac,tot}
+ F_{\rm other}\right]}{{\rm d}h}\,{\rm d}h  = \nonumber\\
&& = F_{\rm ac,tot}(h_2) - F_{\rm ac,tot}(h_1)
+ \Delta F_{\rm other} = \int Q\,{\rm d}h \equiv Q_{\rm tot},
\label{eq:heatintegrated}
\end{eqnarray}
where $h_1=1000$~km, $h_2=1400$~km, and we ignored the contribution of
other energy sources $\Delta F_{\rm other}$. The acoustic fluxes
were calculated using densities and gas pressures from the models
{\it VAL3C} for the LB and {\it VAL3B} for the quiet chromosphere.
A magnetic field inclined by 50\degr~ was assumed in the LB.
The estimated values of acoustic fluxes and radiative losses of
Ca~II + Mg~II are summarised in Table~\ref{tab:energetics}.

\begin{table}
\caption[]{\label{tab:radiativelosses}
Total radiative losses $Q_{\rm tot}$ of Ca II integrated over geometrical
heights 600--2000 km.
}
\centering
\begin{tabular}{lc}
\hline\hline
\noalign{\smallskip}
Model atmosphere & $Q_{\rm tot}$ (Ca II) \\
\hline
\rule{0pt}{3.5mm}VAL3A & 1684~W\,m$^{-2}$ \\
VAL3B & 2345~W\,m$^{-2}$ \\
VAL3C & 3200~W\,m$^{-2}$ \\
\hline
\end{tabular}

\end{table}

\begin{table}
\caption{\label{tab:energetics}
Acoustic power fluxes $F_{\rm ac,tot}$ at geometrical heights
$h_1 = 1000$~km and $h_2 = 1400$~km, their difference
$\Delta F_{\rm ac,tot}$, and an estimate
$Q_{\rm tot}$ of total radiative losses of Ca II and Mg II.
}
\centering
\begin{tabular}{llcccc}
\hline\hline
\noalign{\smallskip}
Region & Model & $F_{\rm ac,tot}(h_1)$ & $F_{\rm ac,tot}(h_2)$ & $\Delta F_{\rm ac,tot}$ & $Q_{\rm tot}$ \\
& & [W\,m$^{-2}$] & [W\,m$^{-2}$] & [W\,m$^{-2}$] & [W\,m$^{-2}$]\\
\hline
\rule{0pt}{3.5mm}Quiet & VAL3B & 470 & 50 & 420 & 3100 \\
LB & VAL3C & 3400 & 400 & 3000 & 4300 \\
\hline
\end{tabular}
\end{table}

Taking into account that the error level in the determination of the total
acoustic flux may be of the order of tens of per cent and the obtained
values are probably underestimated (see Section~\ref{subsec:LBosc}), one
has to look at the numbers more qualitatively. The calculated radiative
cooling in the quiet chromosphere is very probably overestimated -- as we
mentioned, the profile of the Ca~II line corresponds to a model cooler
than {\it VAL3B}, between {\it VAL3A} and {\it VAL3B}. It is evident
that in the quiet chromosphere, the $p$ modes leaking along the small-scale
magnetic fields are not important for the heating of the plasma and the
term $\Delta F_{\rm other}$ dominates the left-hand side of
Equation~(\ref{eq:heatintegrated}).
However, in the chromosphere above the LB, the increased heating by $p$ modes
leaking along the inclined magnetic field practically fully explains the
excess in plasma radiation.  To verify this claim, detailed time-dependent
MHD and radiative-transfer models are needed.

\section{Discussion and conclusions}\label{sec:disc}

We studied the photosphere and chromosphere of a large solar pore
with a granular LB using spectral-imaging observations in the line
Ca~II 854.2~nm and complementary spectropolarimetric observations
in photospheric lines of Fe~I. Dopplergrams mapping chromospheric
LOS velocities at the height of approximately 1000 km were derived
from the Ca~II line core and power maps of acoustic oscillations
were calculated for frequency bands around 1--9 mHz. The spatial
resolution was approximately 0\farcs 4.

We showed that the chromospheric filamentary structure around
the pore, observed in the Ca~II line core and Doppler maps, has all
the important characteristics of a superpenumbra. It is composed of
radially oriented fibrils, which are spatially uncorrelated with
the fibrils seen in the Dopplergrams. The spatial distribution of
time-averaged LOS velocities corresponds to the inverse Evershed
effect. Concentric running waves propagate through the filamentary
structure with the speed of sound and the observed characteristics
of acoustic oscillations correspond to those in a superpenumbra.
Although a strong horizontal magnetic field responsible for the
formation of a penumbra is missing in the pore, the magnetic field
of $B \simeq 1500$~G at the pore's border inclined by 40\degr~
\citep{SobMorEtal:2012} manifests itself in the middle photosphere
by short bright filaments seen in the Ca~II line wings
(Fig.~\ref{fig:exa}b) and its field lines, which return to
opposite-polarity patches dispersed in the photosphere around
the pore, very probably form the filamentary structure with
superpenumbral characteristics in the chromosphere. From this
point of view, the presence of a penumbra is not a necessary
condition for the formation of a superpenumbra.

Horizontal motions of photospheric structures around the pore were
measured using the LCT method at two different heights:
$h\simeq0$~km (continuum images around 621 nm) and $h\simeq250$~km
(images in Ca~II blue wing $-60$~pm), where the reverse granulation
is observed. We found that horizontal flows with typical
centres of diverging motions, corresponding to mesogranules, are
identical at both heights. LCT measurements in the Ca~II line
centre images and Dopplergrams, that is, in the middle chromosphere,
reveal radial motions away from the pore that are located inside the
superpenumbra. Their origin is not clear yet and needs additional
investigation.

Umbral flashes observed in both cores of the pore seem to be
excited by shock waves \citep[cf.][]{Rouppe:2003} that propagate
upwards with a supersonic speed of 20 km~s$^{-1}$.
An enhanced power of acoustic oscillations at frequencies
around 3--5 mHz was observed in the LB inside the pore, in a plage
eastward of the pore, and at the edges of umbral cores.
We assume that these are the regions where acoustic waves leak up
along the inclined magnetic field into the chromosphere, as predicted
by \citet{depontieu2004} and \citet{Cally:2006}.

Special attention was paid to the granular LB that separated the
pore into two umbral cores. Its width measured in magnetic field
maps decreases with increasing height in the atmosphere, which
confirms the magnetic canopy structure \citep{JurLB:2006}.
On the other hand, the width measured in intensity and temperature
maps increases with height as well as the size of LB granules.

In the middle photosphere ($h \simeq 250$ km, Ca~II wings),
reverse granulation is seen around the pore, but not in the LB.
The reverse granulation is explained by adiabatic cooling of
expanding gas in granules, which is only partially cancelled by
radiative heating \citep{Cheung:2007}. In the LB, hot
magnetoconvective plumes at the bottom of the photosphere cannot
expand adiabatically in higher photospheric layers because of
the magnetic field and the radiative heating dominates,
forming small bright granules separated by dark lanes.
The positive correlation between the LB structures in the continuum
and Ca~II line wings indicates that the middle-photosphere
structures are heated by radiation from the low photosphere.
Since the mean free photon path in the photosphere is longer than
1\arcsec ~for $h > 120$~km, the LB observed in the line wings is
broader and its granules are larger than in the continuum due to
the radiation diffusion.

In the middle chromosphere ($h \simeq 1300$ km, Ca~II centre),
the LB is the brightest feature in the pore and it is brighter
by factor of 1.3 than the average intensity in the FOV. Since
the height in the atmosphere is well above the temperature minimum,
radiative heating cannot be expected. The heating by acoustic
waves seems to be a candidate, because the measured acoustic power
flux in the LB is five to seven times higher than that in the quiet
chromosphere. This is because $p$ modes leak along the inclined
magnetic field in the LB and propagate into the chromosphere.
To investigate this possibility, we compared the acoustic power
flux with total radiative losses and found that the acoustic power
flux dissipation in the LB, estimated to 3000 W~m$^{-2}$ as a
minimum, can indeed account quite well for the total radiative
losses (4300 W~m$^{-2}$) computed using the {\it VAL3C} model,
which fits our observed CaII 854.2 nm line profile in the LB.

However, this profile is averaged in time and over the LB area,
which may have certain consequences for the modelling. As is
now well established from radiation-hydrodynamical simulations
of the solar chromosphere \citep[e.g.,][]{CarStein:1998},
semi-empirical models based on time-averaged line profiles
generally overestimate the true mean temperature, especially
in the UV, where the line-intensity averaging is highly nonlinear.
Therefore our computed radiative losses based on this enhanced
empirical temperature and density distribution have to be
considered with caution. It may be that the acoustic waves do not
lead to an enhancement of the mean chromospheric temperature,
but rather cause episodic brightenings within the acoustic shocks,
where the wave energy converts almost instantaneously into
radiation -- this is at least the case of the quiescent
chromosphere in the cell interior \citep{CarStein:1998}.
Therefore, to confirm our conclusions about the wave heating,
detailed time-dependent simulations are needed.

\begin{acknowledgements}
We thank Marta Vantaggiato for her participation in the observations.
This work was supported by the Czech Science Foundation under grants
P209/12/0287 and P209/12/P568 and by the project RVO:67985815
of the Academy of Sciences of the Czech Republic.
The Dunn Solar Telescope is located at the National Solar Observatory,
which is operated by the Association of Universities for Research in
Astronomy, Inc. (AURA), for the National Science Foundation.
IBIS has been built by INAF/Osservatorio Astrofisico di Arcetri with
contributions from the Universities of Firenze and Roma ``Tor Vergata'',
the National Solar Observatory, and the Italian Ministries of Research
(MIUR) and Foreign Affairs (MAE). HINODE is a Japanese mission developed
and launched by ISAS/JAXA, with NAOJ as domestic partner and NASA and
STFC (UK) as international partners. It is operated by these agencies
in cooperation with ESA and NSC (Norway).
\end{acknowledgements}

%
\bibliographystyle{aa} 

\begin{thebibliography}{}
%
\bibitem[Alissandrakis et al.(1988)]{Aliss:1988}
Alissandrakis, C. E., Dialetis, D., Mein, P., Schmieder, B.,
\& Simon, G. 1988, \aap, 201, 339
%
\bibitem[Beckers \& Schr\"oter(1969)]{BeckSchr:1969}
Beckers, J. M., \& Schr\"oter, E. H. 1969, Solar Phys., 10, 384
%
\bibitem[Beckers \& Schultz(1972)]{BeckSchultz:1972}
Beckers, J. M., \& Schultz, R. B. 1972, Solar Phys., 27, 61
%
\bibitem[Beckers \& Tallant(1969)]{BeckTal:1969}
Beckers, J. M., \& Tallant, P. E. 1969, Solar. Phys., 7, 351
%
\bibitem[Bello Gonz\'alez et al.(2009)]{bg2009}
Bello Gonz{\'a}lez, N., Flores Soriano, M., Kneer, F., \& Okunev, O. 2009,
\aap, 508, 941 
%
\bibitem[Berger \& Berdyugina(2003)]{BeBe:2003}
Berger, T. E., \& Berdyugina, S. V. 2003, \apj, 589, L117
%
\bibitem[Bhatnagar et al.(1972)]{Bhatnagar:1972}
Bhatnagar, A., Livingston, W. C., \& Harvey, J. W. 1972, Solar Phys., 27, 80
%
\bibitem[Bogdan \& Judge(2006)]{Bogdan:2006}
Bogdan, T. J., \& Judge, P. G. 2006, Phil. Trans. R. Soc. A, 364, 313
%
\bibitem[Bonet et al.(2005)]{Bonet:2005}
Bonet, J. A., M\'arquez, I., Muller, R., Sobotka, M., \& Roudier, Th.
2005, \aap, 430, 1089
%
\bibitem[Brynildsen et al.(1999)]{Brynild:1999}
Brynildsen, N., Maltby, P., Leifsen, T., et al. 1999, Solar Phys., 191, 129
%
\bibitem[Cally(2006)]{Cally:2006}
Cally, P. S. 2006, Royal Society of London Transactions Series A,
vol. 364, Issue 1839, 333
%
\bibitem[Carlsson \& Stein(1998)]{CarStein:1998}
Carlsson, M., \& Stein, R. F. 1998, in New Eyes to See Inside Sun and Stars,
ed. F.-L. Deubner, J. Christensen-Dalsgaard, and D. Kurtz, IAU Symp. 185, 435
%
\bibitem[Cauzzi et al.(2008)]{Cauzzi:2008}
Cauzzi G., Reardon, K. P., Uitenbroek, H., et al. 2008, \aap, 480, 515
%
\bibitem[Cavallini(2006)]{Cavallini:2006}
Cavallini, F. 2006, Solar Phys., 236, 415
%
\bibitem[Cheung et al.(2007)]{Cheung:2007}
Cheung, M. C. M., Sch\"ussler, M., \& Moreno Insertis, F. 2007, \aap, 461, 1163
%
\bibitem[Christensen-Dalsgaard et al.(1996)]{jcd1996}
Christensen-Dalsgaard, J., Dappen, W., Ajukov, S. V., et al.\ 1996, 
Science, 272, 1286 
%
\bibitem[Christopoulou et al.(2000)]{Christop:2000}
Christopoulou, E. B., Georgakilas, A. A., \& Koutchmy, S. 2000, \aap, 354, 305
%
\bibitem[De Pontieu et al.(2004)]{depontieu2004}
De Pontieu, B., Erd{\'e}lyi, R., \& James, S. P. 2004, \nat, 430, 536 
%
\bibitem[Garcia et al.(2010)]{Garcia:2010}
Garcia, A., Klva\v{n}a, M., \& Sobotka, M. 2010,
Cent. Eur. Astrophys. Bull., 34, 47
%
\bibitem[Giordano et al.(2008)]{giordano:08}
Giordano, S., Berrilli, F., Del Moro, D., \& Penza, V. 2008,
\aap, 489, 747
%
\bibitem[Giovanelli(1972)]{Giovanelli:1972}
Giovanelli, R. G. 1972, Solar Phys., 27, 71
%
\bibitem[Hirzberger(2003)]{hirzberg:03}
Hirzberger, J. 2003, \aap, 405, 331
%
\bibitem[Jefferies et al.(2006)]{jefferies2006}
Jefferies, S.~M., McIntosh, S.~W., Armstrong, J.~D., et al.\ 2006,
\apjl, 648, L151 
%
\bibitem[Jur\v{c}\'{a}k et al.(2006)]{JurLB:2006}
Jur\v{c}\'{a}k, J., Mart\'\i nez Pillet, V., \& Sobotka, M.
2006, \aap, 453, 1079
%
\bibitem[Keil et al.(1999)]{keiletal:1999}
Keil, S. L., Balasubramaniam, K. S., Smaldone, L. A.,
\& Reger, B. 1999, \apj, 510, 422
%
\bibitem[Keppens \& Mart\'\i nez Pillet(1996)]{kepmar:96}
Keppens, R., \& Mart\'\i nez Pillet, V. 1996, \aap, 316, 229
%
\bibitem[Khomenko \& Cally(2012)]{KhoCall:2012}
Khomenko, E., \& Cally, P. S. 2012, \apj, 746, 68
%
\bibitem[Klimchuk(2006)]{Klimchuk:2006}
Klimchuk, J. A. 2006, Solar Phys., 234, 41
%
\bibitem[Kosugi et al.(2007)]{Kosugi:2007}
Kosugi, T., Matsuzaki, K., Sakao, T., et al. 2007, Solar Phys., 243, 3
%
\bibitem[Leka(1997)]{Leka:1997}
Leka, K. D. 1997, \apj, 484, 900
%
\bibitem[Linsky et al.(1970)]{Linsky:1970}
Linsky, J. F., Teske, R. G., \& Wilkinson, C. W. 1970, Solar Phys., 11, 374
%
\bibitem[Lites(1992)]{Lites:1992}
Lites, B. W. 1992, in Sunspots: Theory and Observations, ed.
J. H. Thomas \& N. O. Weiss, NATO ASI Series, C375 (Dordrecht: Kluwer), 261
%
\bibitem[Lites et al.(1991)]{Lites:1991}
Lites, B. W., Bida, T. A., Johannesson, A., \& Scharmer, G. B.
1991, \apj, 373, 683
%
\bibitem[Maltby et al.(1986)]{Maltby:1986}
Maltby, P., Avrett, E. H., Carlsson, M., et al. 1986, \apj, 306, 284
%
\bibitem[November \& Simon(1988)]{novsim:1988}
November, L. J., \& Simon, G. W. 1988, \apj, 333, 427
%
\bibitem[Ortiz et al.(2010)]{Ortizetal:2010}
Ortiz, A., Bellot Rubio, L. R., \& Rouppe van der Voort, L. 2010,
\apj 713, 1282
%
\bibitem[Reardon \& Cavallini(2008)]{ReCa:2008}
Reardon, K. P. \& Cavallini, F. 2008, \aap, 481, 897
%
\bibitem[Reznikova et al.(2012)]{reznikova2011}
Reznikova, V. E., Shibasaki, K., Sych, R. A., \& Nakariakov, V. M. 2012,
\apj, 746, 119 
%
\bibitem[Rieutord et al.(2010)]{Rieutord:2010}
Rieutord, M., Roudier, T., Rincon, F., et al. 2010, \aap, 512, A4
%
\bibitem[Rimmele(1997)]{Rimmele:1997}
Rimmele, T. R. 1997, \apj, 490, 458
%
\bibitem[Rimmele et al.(2004)]{Rimmele:2004}
Rimmele, T. R., Richards, K. Hegwer, S., et al. 2009, in Telescopes
and Instrumentation for Solar Astrophysics, ed. S. Fineschi \& M. A.
Gummin, Proceedings of the SPIE, 5171, 179
%
\bibitem[Rouppe van der Voort et al.(2003)]{Rouppe:2003}
Rouppe van der Voort, L. H. M., Rutten, R. J., S\"utterlin, P.,
et al. 2003, \aap, 403, 277
%
\bibitem[Rucklidge et al.(1995)]{Rucklidge:1995}
Rucklidge, A. M., Schmidt, H. U., \& Weiss, N. O. 1995,
\mnras, 273, 491
%
\bibitem[Ruiz Cobo \& del Toro Iniesta(1992)]{RuizCobo:1992}
Ruiz Cobo, B., \& del Toro Iniesta, J. C. 1992, \apj, 398, 375
%
\bibitem[R\"uedi et al.(1995)]{Ruedi:1995}
R\"uedi, I., Solanki, S. K., \& Livingston, W. 1995, \aap, 302, 543
%
\bibitem[Rybicki \& Hummer(1991)]{RybHum:1991}
Rybicki, G. B., \& Hummer, D. G. 1991, \aap 245, 171
%
\bibitem[Rybicki \& Hummer(1992)]{RybHum:1992}
Rybicki, G. B., \& Hummer, D. G. 1992, \aap 262, 209
%
\bibitem[Scherrer et al.(2012)]{Scherrer:2012}
Scherrer, P. H., Schou, J., Bush, R. I., et al. 2012, Solar Phys.,
275, 207
%
\bibitem[Schou et al.(2012)]{Schou:2012}
Schou, J., Scherrer, P. H., Bush, R. I., et al. 2012, Solar Phys.,
275, 229
%
\bibitem[Shimizu(2012)]{Shimi:2012}
Shimizu, T. 2012, in Hinode-3: The 3rd Hinode Science Meeting,
ed. T. Sekii, T. Watanabe, and T. Sakurai, ASP Conference series,
Vol. 454, 177
%
\bibitem[Simon \& Weiss(1970)]{SimonWeiss:1970}
Simon, G. W., \& Weiss, N. O. 1970, Solar Phys., 13, 85
%
\bibitem[Sobotka(1999)]{sobotka:1999}
Sobotka, M. 1999, in Motions in the Solar Atmosphere, ed. A. Hanslmeier
and M. Messerotti, Dordrecht: Kluwer, 71
%
\bibitem[Sobotka et al.(1997)]{sbs:1997}
Sobotka, M., Brandt, P. N., \& Simon, G. W. 1997, \aap, 328, 682
%
\bibitem[Sobotka et al.(2012)]{SobMorEtal:2012}
Sobotka, M., Del Moro, D., Jur\v c\'ak, J., \& Berrilli, F. 2012,
\aap, 537, A85
%
\bibitem[Sobotka \& Jur\v c\'ak(2009)]{sobjur:09}
Sobotka, M., \& Jur\v c\'ak, J. 2009, \apj, 694, 1080
%
\bibitem[Sobotka et al.(1999)]{sobetal:99}
Sobotka, M., V\'azquez, M., Bonet, J. A., Hanslmeier, A.,
\& Hirzberger, J. 1999, \apj, 511, 436
%
\bibitem[Stangalini et al.(2011)]{Stangal:2011}
Stangalini, M., Del Moro, D., Berrilli, F., \& Jefferies, S. M., 2011,
\aap, 534, A65
%
\bibitem[Stangalini et al.(2012)]{Stangal:2012}
Stangalini, M., Giannattasio, F., Del Moro, D., \& Berrilli, F. 2012,
\aap, 539, L4
%
\bibitem[S\"utterlin(1998)]{Suetter:98}
S\"utterlin, P. 1998, \aap, 333, 305
%
\bibitem[Thomas \& Weiss (2008)]{TW:2008}
Thomas, J.H., \& Weiss, N.O. 2008, Sunspots and Starspots,
Cambridge University Press
%
\bibitem[Tsiropoula et al.(1996)]{Tsiro:1996}
Tsiropoula, G., Alissandrakis, C. E., Dialetis, D., \& Mein, P. 1996,
Solar Phys., 167, 79
%
\bibitem[Tsuneta et al.(2008)]{Tsuneta:2008}
Tsuneta,  S., Ichimoto, K., Katsukawa, Y., et al. 2008, Solar Phys., 249, 167
%
\bibitem[Tziotziou et al.(2006)]{Tzio:2006}
Tziotziou, K., Tsiropoula, G., Mein, N., \& Mein, P. 2006, \aap, 456, 689
%
\bibitem[Tziotziou et al.(2007)]{tziotzio2007}
Tziotziou, K., Tsiropoula, G., Mein, N., \& Mein, P. 2007, \aap, 463, 1153 
%
\bibitem[Uitenbroek(1989)]{Uitenbroek:1989}
Uitenbroek, H. 1989, \aap, 213, 360
%
\bibitem[van Noort et al.(2006)]{Noort:2006}
van Noort, M., Rouppe van der Voort, L., \& L\"ofdahl, M. 2006,
in Solar MHD Theory and Observations, ed. J. Leibacher, R. F. Stein,
and H. Uitenbroek, ASP Conference Series, Vol. 354, 55
%
\bibitem[Verma et al.(2013)]{Verma:2013}
Verma, M., Steffen, M., Denker, C. 2013, arXiv:1305.6033
%
\bibitem[Vernazza et al.(1981)]{VAL3:1981}
Vernazza, J. E., Avrett, E. H., \& Loeser, R. 1981, \apjs, 45, 635
%
\bibitem[Vissers \& Rouppe van der Voort(2012)]{Vissers:2012}
Vissers, G., \& Rouppe van der Voort, L. 2012, \apj, 750, 22
%
\bibitem[Viticchi\'{e} et al.(2010)]{viticchie2010}
Viticchi\'{e}, B., Del Moro, D., Criscuoli, S., \& Berrilli, F.
2010 \apj, 723, 787
%
\bibitem[Zirin \& Stein(1972)]{Zirin:1972}
Zirin, H., \& Stein, A. 1972, \apj, 178, L85
%
\end{thebibliography}

\end{document}